# Towards Refactoring of DMARF and GIPSY Case Studies – A Team 5 SOEN6471-S14 Project Report

Pavan Kumar Polu
ENCS
Concordia University
Montreal, Canada
p_polu@encs.concordia.ca

Amjad Al Najjar
ENCS
Concordia University
Montreal, Canada
a_alna@encs.concordia.ca

Biswajit Banik
ENCS
Concordia University
Montreal, Canada
bi_banik@encs.concordia.ca

Ajay Sujit Kumar
ENCS
Concordia University
Montreal, Canada
aj_vijay@encs.concordia.ca

Gustavo Pereira
ENCS
Concordia University
Montreal, Canada
gu_perei@encs.concordia.ca

Prince Japhlet
ENCS
Concordia University
Montreal, Canada
p_stephe@encs.concordia.ca

Bhanu Prakash R.
ENCS
Concordia University
Montreal, Canada
b_bamine@encs.concordia.ca

Sabari Krishna Raparla
ENCS
Concordia University
Montreal, Canada
s_raparl@encs.concordia.ca

*Abstract:* **This paper presents an analysis of the architectural design of two distributed open source systems (OSS) developed in Java: Distributed Modular Audio Recognition Framework (DMARF) and General Intensional Programming System (GIPSY). The research starts with a background study of these frameworks to determine their overall architectures. Afterwards, we identify the actors and stakeholders and draft a domain model for each framework. Next, we evaluated and proposed a fused DMARF over GIPSY Run-time Architecture (DoGRTA) as a domain concept. Later on, the team extracted and studied the actual class diagrams and determined classes of interest. Next, we identified design patterns that were present within the code of each framework. Finally, code smells in the source code were detected using popular tools and a selected number of those identified smells were refactored using established techniques and implemented in the final source code. Tests were written and ran prior and after the refactoring to check for any behavioral changes.**

## I. INTRODUCTION

It is important to generate, evaluate and maintain a well-structured and documented system architecture to enable a stable and maintainable software system. This paper is an analysis of the architectural design of two distributed open source systems (OSS) developed in Java: Distributed Modular Audio Recognition Framework (DMARF) and General Intensional Programming System (GIPSY). The team led a research to understand these architectures and steered this analysis deep enough to learn about their design patterns and design decisions. Some tools like CodePro Analytix, SonarQube and InCode were also used in source code analysis and helped in generating metrics report. A conceptual understanding of the case studies domain was done by designing UML Domain diagram using the knowledge gained from the studied articles, followed by a proposed conceptual merger of DMARF and GIPSY Run-time Architecture (DoGRTA).

We followed this up with an actual study of the Classes in GIPSY and DMARF, comparing these resulting class diagrams with the respective conceptual domain models. The similarities and differences were noted down. Each team member was tasked with the responsibility of identifying a unique pattern in the software systems. The understanding of architecture and patterns helped us to identify the flaws in the current code which were noted down as code smells. Tools like JDeodorant, PMD, and CodePro helped with the identification of code smells and ObjectAid in visualizing classes and relationships. Few refactoring tactics were proposed, discussed and implemented as part of refactoring the source code. Significant smells with appropriate refactoring techniques were chosen to be implemented in the final milestone. The refactored code was verified with test cases written to check for variations in behavior. The following sections will explain these steps in details.

## II. BACKGROUND

The research started with the study of eight papers related to these two frameworks, DMARF and GIPSY. The team members could read about different aspects of each framework. A conceptual understanding and domain knowledge was built on the analysis of each paper and the overall architectures could be recognized. The result was the development of a UML Domain diagram using the knowledge gained in these articles, followed by a proposed conceptual linking of DMARF and GIPSY as distributed systems. We also used some tools to extract information from the source code of each frameworks and ObjectAid tool to extract the actual class diagrams. We could analyze classes, relationships and their responsibilities. The information extracted from these papers is showed in the chapter for each framework.

### A. OSS CASE STUDIES

#### 1) DMARF: Distributed Modular Audio Recognition Framework

The Modular Audio Recognition Framework (MARF) is an open-source framework built in Java used to recognize and identify audio and classify such identifications using pattern recognition algorithms and natural language processing (NLP). [1, 2] The original implementation of MARF of this framework was sequential with little to no concurrency. This design added some limitations and made it difficult to scale and to accommodate large amount of data [1]. Thus was born DMARF - Distributed Modular Audio Recognition Framework. Figure 1 shows DMARF overall architecture.

DMARF is a modified version of classical MARF where the pipeline stages play a role of distributive nodes. The MARF pipeline is composed of four core stages:

1. Sample loading
2. Pre-processing
3. Feature extraction
4. Training/Classification.

DMARF separated these stages and aimed to offer them as services over a distributed network therefore thus offloading the bulk of data crunching to dedicated high-performance machines [1]. A second motivation is to implement disaster recovery and replication techniques featured in a distributed system environment. The distributed design of the MARF pipeline is presented in Figure 2 showing the different front-end levels, back-end modules and the communication flow.

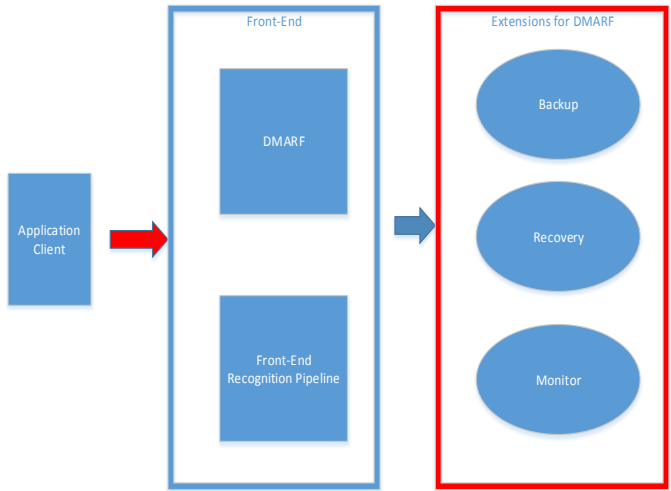

Figure 1: Simplified DMARF Overall architecture

DMARF was designed with the following architectural strategies in mind: Platform-Independence, Database-Independent API, Communication Technology Independence, Reasonable Efficiency, Simplicity and Maintainability, Architectural Consistency and Separation of Concern [1]. As a project used also for learning and research, DMARF was implemented with three different distributed technologies: CORBA, RMI and Web services.

This architectural design shows the possibility in providing services to clients that have low computational power where the resources consumed by services can rely on the servers.

DMARF's well-structured architecture allowed people to imagine creative ways to improve or evolve on it. We've examined several papers that apply such possibilities.

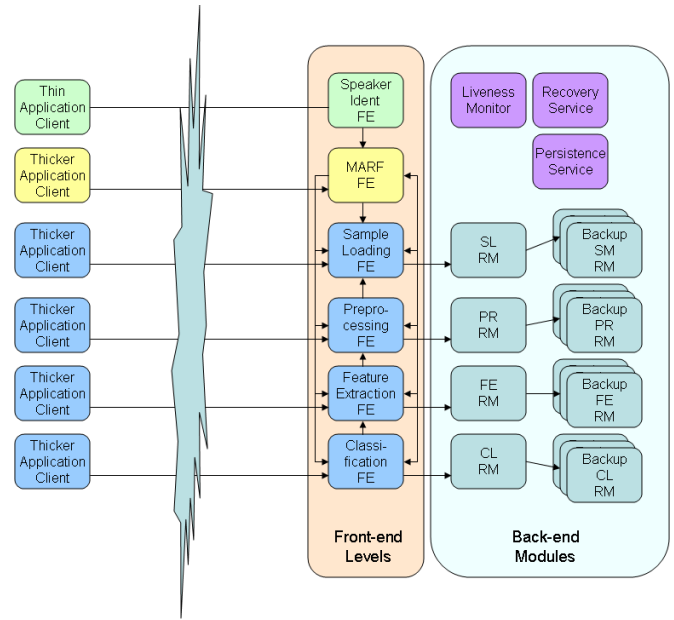

Figure 2: The Distributed MARF Pipeline [3]

One repeated concept is the notion of "autonomic" DMARF (ADMARF) which refers to self-managing characteristics, thus eliminating (or at least significantly reducing) the need to user input. The goal is to allow the system to run itself.

In [4], the focus is on how self-healing mechanism of DMARF works using Autonomic System Specification Language (ASSL) [4]. DMARF basically depends on three autonomic requirements like self-healing, self- protection and self-optimization. The idea was to strategically devise algorithms in ASSL for the pipelined stages of DMARF so that each stage can execute reliably for long periods of time. [4] Therefore, if we detect that a particular stage has gone offline or fails to respond, then the algorithm kicks in and either: replaces/recovers the irresponsive node or reroute the pipeline via an alternate node allowing the system to seamlessly progress. [4] ASSL provides DMARF an autonomic middleware enabling it to perform in autonomous systems reducing human intervention. Self- healing is accomplished by examining the properties of DMARF. ADMARF monitors its

runtime performance if the performance is degraded then DMARF notifies to start the self-healing. [4]

In [5], this autonomic notion is abstracted further to pattern recognition systems. ASSL was used again to develop such properties for DMARF. This suggested approach will help create a pattern-recognition pipeline. Autonomic system helps reduce the necessary workload for maintaining complex system by transforming them to self-managing systems. Pattern recognition is a widely accepted approach for recognizing shapes, images, voices, sounds, etc. ASSL provides many properties for DMARF such as self-healing, self-optimization and self-protection, all part of self-management idea. These properties generated ASSL framework in the form of special wrapper Java code that provides an autonomic layer implementing the DMARF's autonomic properties.

The vision and metaphor of Autonomic Computing [8, 15, 16] is applied to the self-regulation principles for complexity hiding. The main idea behind this approach is that software systems will act automatically like as a human body's nervous system does. ASSL approaches the problem of formal specification and code generation of autonomic systems (ASs) within a framework and helps to create an operational Java application skeleton. ASSL also considers autonomic systems (ASs) as composed of autonomic elements (AEs) communicating over interaction protocols. ASSL is defined as formalization of tiers, where it provides multi-tier specification model. This model is designed and composed of infrastructure elements that are needed by an AS. The AS specifies itself in terms of **Service-Level Objectives (AS SLOs)**, self-management policies, architecture topology, actions, events, and metrics. The AS SLOs are a high-level form of behavioral specification which helps developers to establish system objectives (e.g., performance). The main target of ASSL [2, 3, 4] is to feature AC systems and help developers to distinguish between AC features and system features. In case of self-protection systems based on DMARF where each node has to prove its identity to other nodes those which are participating in the pipeline. This is the approach by which we can ensure that the distributed data is processed correctly.

Another concept that is presented in several papers is the notion of interoperability which aims to allow DMARF's services to be accessible in multiple ways (RMI, CORBA, RPC, Web services).

In [6], the web services handling the Remote Procedure Call (RPC) are explored. DMARF is implemented in Java with modules for invoking its services in RMI and CORBA modules. They can replace each other and aren't met with flexibility, which needs support from the clients of Java or CORBA that are not in MARF modules. This can be overcome by the introduction of web services and make it compatible with RMI and COBRA services. The result is that the DMARF nodes are more interoperable over online network or simply HTTP restricted environments. The general MARF applications in the desktop systems are the audio recording, enabling high volume processing, textual or imagery data (this holds for pattern recognition and helps in biometric forensic analysis). DMARF composes a distributed process with web services and offers greater availability over the Internet [6].

In [7], we are presented with a way to manage DMARF's services over a network as a whole. DMARF's stand-alone components may receive requests from RMI, XML-RPC, CORBA and TCP connections and do not natively understand the popular and familiar simple network management protocol (SNMP) [7]. Thus, the goal was to prototype an extension to DMARF to support SNMP. Therefore, the authors looked to add a proxy pattern to act as an interface between SNMP and DMARF. [7]. SNMP itself does not define what information a managed system presents; rather it is defined by management information bases (MIB). Therefore, it is necessary to define MIBs for MARF Services. [7] Figure 3 presents a preface of the management architecture for MARF application and services.

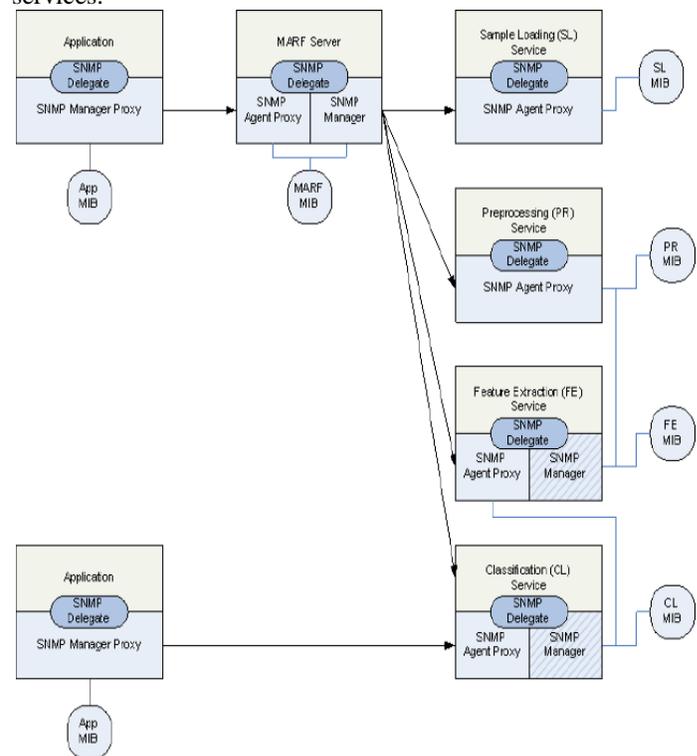

Figure 3 : MARF-Manager-Agent Architecture [7]

DMARF offers a number of services which each will need to be defined in MIB so that they can be managed remotely over a network as a whole. The authors used the notation defined by Structure of Management Information (SMI), a subset of Abstract Syntax Notation (ANS.1) to describe the MIBs. For each service, they had to introduce MIB modules and associated parameters or attributes. [7]

An example of the SMI tree for the sample loading is shown in Figure 4. The other services were somewhat similar, each representing the attributes and data it needed to carry.

Two type of proxy agents were produced using AdventNet's tool: one to talk to MARF's API and the agent and the other talks to manager and agents by using SNMP. [7] No GUI was produced but AdventNet's MIB Browser was employed to test their work. [7] Managing DMARF over a well-adapted network protocol such a SNMP presents some practical opportunities in the real world. Over a distributed network such police agencies spread out across a country, they may be able to identify speakers across all jurisdictions for a recorded phone conversation [7]. However, such agencies would need to feel confident that conversations analyzed were stored safely and security is guaranteed.

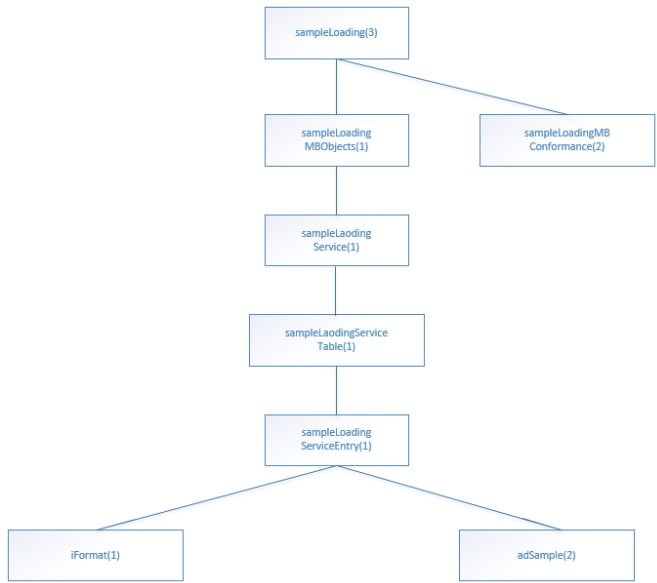

Figure 4: Preliminary MARF Sample Loading Service MIB [8]

This is the topic of the next case study. In [24], tightening security for distributed computing systems is explored. We analyzed a proposed solution using JDSF (Java Data Security Framework) framework to address security aspects of distributed systems such as injection, data alteration, malicious code injection, etc. It is understood that this layer of security may present a significant overhead that can impact performance for DMARF which is expected to process data in a timely manner. Therefore, a configurable option to turn on or off was suggested when communicating through public channels. JDSF design provides an abstraction of the common essential cryptographic primitives for Java open source implementations of the cryptographic algorithms for encryption, hashing, digital signatures, etc. Confidentiality is not given importance in both systems as the design concentrated on correctness and accuracy of the computation and stored results [24]. But application of JDSF's confidentiality sub framework would resolve the confidentiality concerns in both systems. As the case study systems are developed in Java, integration in to security layer is insignificant. So, to provide integration JDSF calls will be injected on the computing and generating nodes that are invoked before the data exits or enter the system. However, this structure assumes that key set up and exchange already happened. Data authentication which relates on integrity and correctness of scientific computation is also important in both systems, because there is a chance of receiving data from an untrusted source that can violate integrity checks and may poison cached results in data store. So, to prevent that JDSF authentication sub framework is used here as similarly used in confidentiality and integrity aspects. Generally, any distributed system should contain the property of redundancy to provide its complete availability of services. But in this case, due to some regular network problems that are connecting participating nodes, malicious code obstruct systems from providing complete availability. In distributed systems it is very hard to guarantee the availability even using JDSF [24]. JDSF's sub frameworks helps in solving aspects like confidentiality, integrity and authentication but might still have some limitations in achieving availability [24].

One other idea brought up in several papers is the notion of self-forensics. Self-forensics is the concept of retrieving information, deducting and reconstructing incidents automatically with an option to be self-diagnostic. The modules are capable of making complex decisions based on evidence. Forensic computing is primarily concerned with computer crime investigations, while Forensic LUCID is the language for self-forensics of computer crime incidents. DMARF functions over the network or as a library function in application and provides an extensive data gathering and coverage as it involves multiple complex communication between modules and access to remote modules. There could be multiple pipelines established in a DMARF network during its computing. Once a self-forensic module is designed with the necessary requirements, it could be used to log information in Forensic LUCID format of the system in question. These logs should be accurate enough to base decisions and diagnostics on the system incidents. Such a system would be useful in cybercrime investigations and analysis and response to system incidents like system failures. With study over time, if a contextual data base is established, it would be easy to link causes to effects and findings can be shared globally.

To reiterated, ASSL is a framework to formally specify and develop self-management features for autonomic systems [25]. Therefore, by extending the framework to include an autonomic property of self-forensics adding, this would allow the quick completion and validation of experiments and their results. Self-forensics autonomic property (SFAP) has been added to ASSL toolkit to generate Java-based object oriented intensional programming language (JOOIP) [26]. The ASSL framework [27] uses the specifications of autonomic system properties [28, 29] as the input, it performs a formal semantics and syntax checks. It then generates a set of classes and interfaces for the specification if the check passes. As we mentioned, ASSL framework works using AS to specify service level objects (SLOs) and defines the protocols for communication between autonomous systems and the architecture of autonomic element for the whole system. The formal modeling, model checking and specification [30] have been added to number of systems like DMARF [31] and GIPSY [32].To implement SFAP, the notion of self-forensics is added to the autonomous system and the autonomic element. There are 2 major parts:

1. The lexical analyzer, parser and semantic checker of ASSL are added with syntax and semantical support.
2. To translate the forensic events code generator for JOOIP and Forensic LUCID are added [33].

To increase or decrease the Forensic LUCID events the ASSL's managed element (ME) specification of AE is used on the software system. The JOOIP [33] code along with the Forensic LUCID fragments are generated by the ASSL toolset, these are sent to the hybrid compiler of GIPSY (GIPC) which will compile and link them as executable code in the GEE engine resources (GEER). The evaluation of this code can then be done using either traditional model of GEE, AspectJ based model or problematic model checking using backend PRISM. The preliminary analysis of the requirements to implement the autonomic property self-forensics in the ASSL toolset along with the process of implementing the self-forensics has been discussed in the case study. However the actual implementation of the property on to systems like DMARF and GIPSY [31, 32] need further analysis.

### B.  *GIPSY - General Intensional Programming System*

Intensional programming allows expressions involved in the program to be defined in a number of dimensions earning the name as a multi-dimensional programming [15]. The General Intensional Programming System (GIPSY) is an open-source framework that aims to provide a platform for the investigation on the intensional and hybrid intentional-imperative programming. [16] This framework is suitable for LUCID programming languages which is a form of intensional programming. In Lucid, the intensional programming is of declarative programs where all the identifiers are defined as expressions using other identifiers and algebra [23].

Expressions written in all LUCID dialects are corresponding to Higher-Order Intensional Logic (HOIL) expressions which can alter based on the context of their evaluation, range and given set of operators. HOIL brings together functional programming and intensional logic. GIPSY provides a platform reasoning these HOIL expressions in a similar way Forensic LUCID reasons about cybercrime incidents. [22]

The overall architecture of GIPSY is represented in the figure 5.

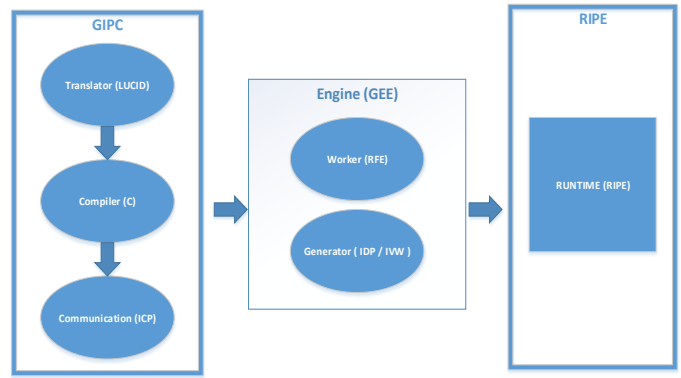

Figure 5: GIPSY Overall architecture

The GIPSY framework is developed in Java and is modular as it is flexible meaning that it is made up of replaceable components with very few conditions. [17, 18] GIPSY is geared towards certain common goals as ensemble below:

- Flexibility – the framework allows compile and run time changes during execution.
- Modularity – the framework is made up of replaceable components that are easily replaced with very few conditions.
- Generality – it is designed in a way to be generic and language independent.
- Adaptability – It is adaptable to different requirements, for example changing the Demand Migration Framework (DMF).
- Efficiency – Improved efficiency using a distributive computing.
- Scalability - the framework design is open for development.

GIPSY's framework is divided into three components: General Intensional Programming Language Compiler (GIPC), Generic Eduction Engine Resources (GEER), and Run-time Interactive Programming Environment (RIPE) [18, 20].

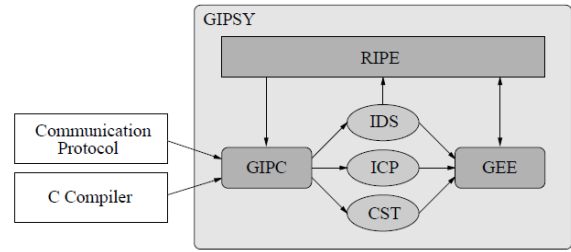

Figure 6: GIPSY Architecture [20]

### 1)  *General Intensional Programming Language Compiler (GIPC):*

The compilation of GIPSY takes place in two stages. First the translation of GIPSY program to C and then the C program is compiled. The source code of GIPSY is subdivided into two parts: LUCID part and Sequential part. The GIPC translates the LUCID part of the program again into two parts

IDS and ICP. The Intensional Data Dependency Structure (IDS) provides the dependency of variables in the LUCID part and Intensional Communication Procedures (ICP) are the data communication procedures used. The C compiler translates the sequential functions of GIPSY to C sequential threads (CST).

### 2) General Eduction Engine (GEE):

A demand-driven model computation is used by GIPSY, i.e. a computation takes place only if there is an explicit demand for it. A procedure call is generated for every demand which can be calculated locally or remotely, the result is stored in a warehouse. This result will be retrieved instead of computing once again if another demand for the same computation is generated. A generator-worker execution architecture is followed by GIPSY. The generator interprets the IDC generated by GIPC. The generator evaluates low-charge ripe sequential threads locally and high-charge ripe sequential threads remotely.

### 3) Run-time Interactive Programming Environment :

The LUCID part of GIPSY can be visualized as a dataflow diagram using RIPE. RIPE provides users with visual run time programming environment with which they can communicate by dynamic inspection of IVW, the input or output channels can be changed, sequential threads can be recompiled, the communication protocol can be changed and also users can change parts of GIPSY.

GIPSY architecture is a multi-tier distributed architecture that has three basic entities and four distinct tiers. The tables below show the breakdown of the basic entities and tiers of the GIPSY runtime system. [19]

| Entities |
| --- |
| GIPSY Tier |
| GIPSY Node |
| GIPSY Instance |

Table 1: GIPSY Entities [19]

| Tiers |
| --- |
| Demand Store Tier (DST) |
| Demand Generator Tier (DGT) |
| Demand Worker Tier (DWT) |
| General Manager Tier (GMT) |

Table 2 : GIPSY Tiers [19]

The execution of GIPSY programs is divided into three different tasks linked to distinct tiers. Communication between the tiers is done using demands. Thus the architecture is described as "demand-driven". [22] There are several kinds of demand types like intensional, procedural, resource and system. Each demand has a universally unique identifier. The Dispatcher class in the system uses the universally unique identifier to identify the kinds of demands that has to be computed in the GIPSY network. For the Demand Migration System (DMS) this identifier remains local while for GIPSY tiers, the identifier remains invisible. Another type of identifier is Demand signature identifier. Here, all the demands that are generated from the same signature will create the same demand signature identifier. The benefit of this approach is that all the same demands after computation can be placed in the DST and their results are easy to extract because there are no need of computation as it follows the principle of dynamic computing. GIPSY was designed with the aim to support a generic and independent collection of Demand Migration System (DMS). It helps to implement DMF. The DMF works to store and communicate information for particular technology. Jini DMS creates a solution based on Jini and Javascript. However, Jini is basically used for design and implementation of the Transport agents (TA) and for Java spaces of the demand store. The JMS –DMS is applied to the DMF framework which is based on Java Messaging Service (JMS). [16, 21]

The flexibility of GIPSY runtime system meant there were a lot of configurable components that were managed primarily via a command-line interface. This was confusing as the user had to memorize commands, syntax and remember the IDs of the nodes and tiers. This presented a big risk as there was no validation in place. This prompted for proposed solution of an interactive graph-based GUI that allowed the user to directly interact with the distributed GISPY runtime system using icons and menu with the addition of data validation to ensure correctness [19]. In the background, the user's clicks and selections were translated into commands that he previously had to type. Another source of potential mistakes was the lengthy configuration files. In the presented solution, these were improved to show only data relevant to the context and the input was validated with corresponding logs and error messages displayed [19].

GIPSY Architecture supports Intensional Cyber forensics which is an extension of Intensional Programming. The idea behind computer forensics is to find or retrieve information related to an event, predominantly crime from computer storages and log files. The information thus retrieved would be used to analyze and reconstruct the event without bias [15]. Interestingly, this forensic analysis is of two types: live and dead analysis. Dead analysis is analysing data and logs after the event has occurred, while live analysis is concerned with dealing with the details of the event as it happens [18].

GIPSY renders to distributed multi-tiered architecture which permits for high scalability but does not have self-management capabilities. This motivated the authors to explore the notion of autonomic for the GIPSY framework just as it was explored for DMARF. Autonomic GIPSY's (AGIPSY) architecture is similar to GIPSY's with the addition of Autonomic Computing (AC) concept which makes complex computing system capable of self-managing itself to an extent. This was achieved based on multiple interacting autonomic GIPSY nodes (GNs) with the support of ASSL. AGIPSY provides the system with goal-driven self-protection, self-healing, self-optimization, and self-configuration characteristics which are necessary for self-management. This model helps to design the architecture of single GNs together with their Service-Level Objectives (SLOs) and self-management policies, the AGIPSY interaction protocol, and the global AGIPSY architectural view and behavior. [17]

## III. METRICS

To compute estimates of GIPSY and MARF, we used a couple of tools to ensure correctness of the reported numbers. For Java files, we used SonarQube [Figure 9] [Figure 12] and we performed a manual search for "*.java" within the project folder. For number of class, methods and lines of code, we looked at three different reports from SonarQube [Figure 9] [Figure 12], CodePro [Figure 8] [Figure 11] and InCode [Figure 7] [Figure 11]. There was a slight discrepancy in the numbers given by InCode, so we took the count reported by the majority.

Estimations were done on the entire code for GIPSY. Estimations for the case study MARF were done only for the specific branch 'DISTRIBUTED_MARF_0_3_0_INTEGRATION'.The results are shown in the below table.

|  | GIPSY | MARF |
|---|---|---|
| **Java Files** | 602 | 1024 |
| **Classes** | 665 | 1054 |
| **Methods** | 5680 | 6305 |
| **Lines of Code** | 104073 | 77297 |

Table 3: Metrics Results

## IV. SUMMARY

The papers analyzed by this article provides basic conceptual model for these two OSS overall architectures: DMARF and GIPSY. Demand-driven GIPSY and pipelined DMARF are distributed frameworks built in a modular manner as to simplify their evolution. Autonomy and having both frameworks work together seems to be the current direction researched.

DMARF is a modified version of classical MARF where the pipeline stages play a role of distributive nodes. This design provides low computational power where the processing could be relayed to run on dedicated high-performance machines. DMARF was implemented with CORBA, RMI and Web services as distributed technologies. Disaster recovery and replication requirements are well satisfied with services provided by back-end modules. ASSL provides DMARF an autonomic middleware enabling it to perform in autonomous systems reducing human intervention.

GIPSY as a conceptual model provides platform for the investigation on the intensional and hybrid intensional-imperative programming. This is provided in a general, adaptive and efficient manner. GIPSY distributed multi-tiered modular architecture deliveries high scalability. AGIPSY is built with ASSL to define the protocols for communication, self-management policies and increasing or decreasing the Forensic LUCID events.

## V. REQUIREMENTS AND DESIGN SPECIFICATIONS

### A. Personas, Actors, and Stakeholders for DMARF

#### 1) Primary Persona:

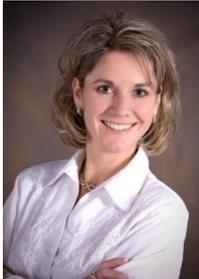

**Alice Desjardins**
30 years old
IT Team Leader and PhD Researcher

Attributes:
Female
Pursued Master's degree in Computer Science
Pursued Bachelor's degree in Information Technology

Alice Desjardins is an IT Development Team Leader in one of the biggest software companies in Montreal. She is 30 years old. She is also a software engineer PhD student leading a research project in Concordia University. For this project, she will need an audio recognition framework for speech recognition that uses natural language processing (NLP) algorithms. She will need a framework that provides means to load the audio and pre-process under a set of desired. It should also allow multiple loads (batch) that identifies the speaker of loaded sample. She would love to use this audio recognition framework in her project in a distributed way. Since the budget for her research project is very small, the hardware capacity of main project server is small, and the resources of this server cannot be consumed by another framework. Using this framework as a service over the network will be helpful for her, since she will need to process a large amount of data. She would love that this framework provides a backup service with a disaster recovery feature.

*2) Actors:*

**User:** a human actor who uses any stages of DMARF System as a frame work through a GUI.

**Application Client:** a non-human actor who uses any stages of DMARF System as a framework invoking DMARF services.

**External Database Server**: a non-human actor that storages all computational data, backup and disaster recovery data from DMARF.

*3) Stakeholders:*

**Users (Clients):** Any person that make use of any DMARF service. This also includes anyone that makes use of the collection of DMARF algorithms or uses as an open source research platform.

**Contributors:** As an open source project many external developers and contributors supports and helps developing and improving the project.

**Developer Team (Project Leaders):** A group of members, who develops, owns and maintains the project. As an open source project, any change by external developers other than development team should be approved by these stakeholders.

B. *Personas, Actors, and Stakeholders for GIPSY*

*1) Persona:*

**Bernard Raj**
25 year old
Software Engineer

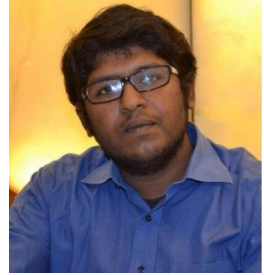

Attributes:
Male
Pursued Master's degree in Computer Science
Pursued Bachelor's degree in Information Technology
Attended many workshops on system application development

Bernard is a Software Engineer that works in a software company. He has good general computer knowledge. He works with some scientific software applications. Sometimes, he struggles with the limitations in these applications. He is looking for an intensional programming language that might help overcome these limitations. He is not that familiar with intensional programming languages or what is required to solve issues within these applications. He prefers to use a graphic user interface where he can fill the necessary fields. He also expects visualize the output in the form of a graph instead of the program code itself. The speech simulated output will be more helpful for him in understanding the problem and solution domain. He needs an intelligent interface that fits for him. As he uses the applications for scientific calculative purpose, he requires the programming system to be more effective, efficient and fault tolerant.

*2) Actors:*

**User:** A typical user can be of any individual who uses GIPSY system. The user can get benefitted with the intensional programming language system applying it to the conflicts that arises with the respective applications.

**External Application**: Any external application that wants to use the part or the whole GIPSY framework. (Ex: DMARF over GIPSY)

*3) Stakeholders:*

**Developers:** The development team of the GIPSY framework, who will support, implement and maintain the system.

**Student:** A student studies the GIPSY system in order to fulfill the project in the grad school. The student acquires knowledge over the system which makes the GIPSY to be a wider range usage in the future.

**Researcher:** The researcher can be of a personality who explores or uses the GIPSY system in a bigger scale for the thesis purpose to excel in the project oriented analysis. The GIPSY system can be requested for any extra feature depends on the benefits to the system and the effective use to the researcher.

C. *Use Cases:*

*1) DMARF*

| Use Case ID: UC1 | Speaker Identification |
|---|---|
| Primary Actor: DMARF User | |
| Stakeholders and Interests: DMARF User: Wants to identify the speaker for the audio file. | |
| Preconditions: None | |
| Preconditions: None | |
| Main Success Scenario: <br><br> 1. The user indicates that he/she wants to load an audio file of a **speaker** he/she wishes to identify. <br><br> 2. The system loads the **sample** and stores the audio file. <br><br> 3. The system preprocesses and normalizes the data and prepares it for **feature extraction**. <br><br> 4. The system extracts **features** from the data to identify the **speaker**. <br><br> 5. The system classifies the readings. <br><br> 6. The system recognizes the **speaker** informs the user of his or her identity. | |
| Extensions: 5b. The system is unable to recognize the user and instead stores the data to train the system. | |

*2) GIPSY*

| Use Case ID: UC2 | Setup GIPSY network |
|---|---|
| Primary Actor: GIPSY user | |
| Level: User Goal | |
| Stakeholders and Interests: GIPSY User: Wants to setup a GIPSY network. | |
| Preconditions: Appropriate configuration files for each tier type are created. | |
| Post conditions: A GIPSY network is started. | |
| Main Success Scenario: <br><br> 1. The user indicates that he/she wants to create a new **GIPSY** | |

**node**.

2. The system starts the bootstrap process and instantiates a **GMT tier** with the setting and properties from the configuration file.

3. The system creates a **GIPSY node** and registers the node and a registration DST is allocated.

4. The user indicates that he/she wants to allocate a **DST** to the registered node.

5. The system sends a request to the **GMT** with the node identifier where a **DST** instance will be allocated.

6. The user indicates that he/she wants to allocate a **DWT** to the registered node.

7. The system sends a request to the **GMT** with the node identifier where a **DWT** instance will be allocated.

8. The system indicates to the GIPSY network is ready to receive demands.

Extensions:

2a. If the configuration file contains invalid properties, notify the user.

D. *Domain Model UML Diagrams*

*1)* *DMARF*

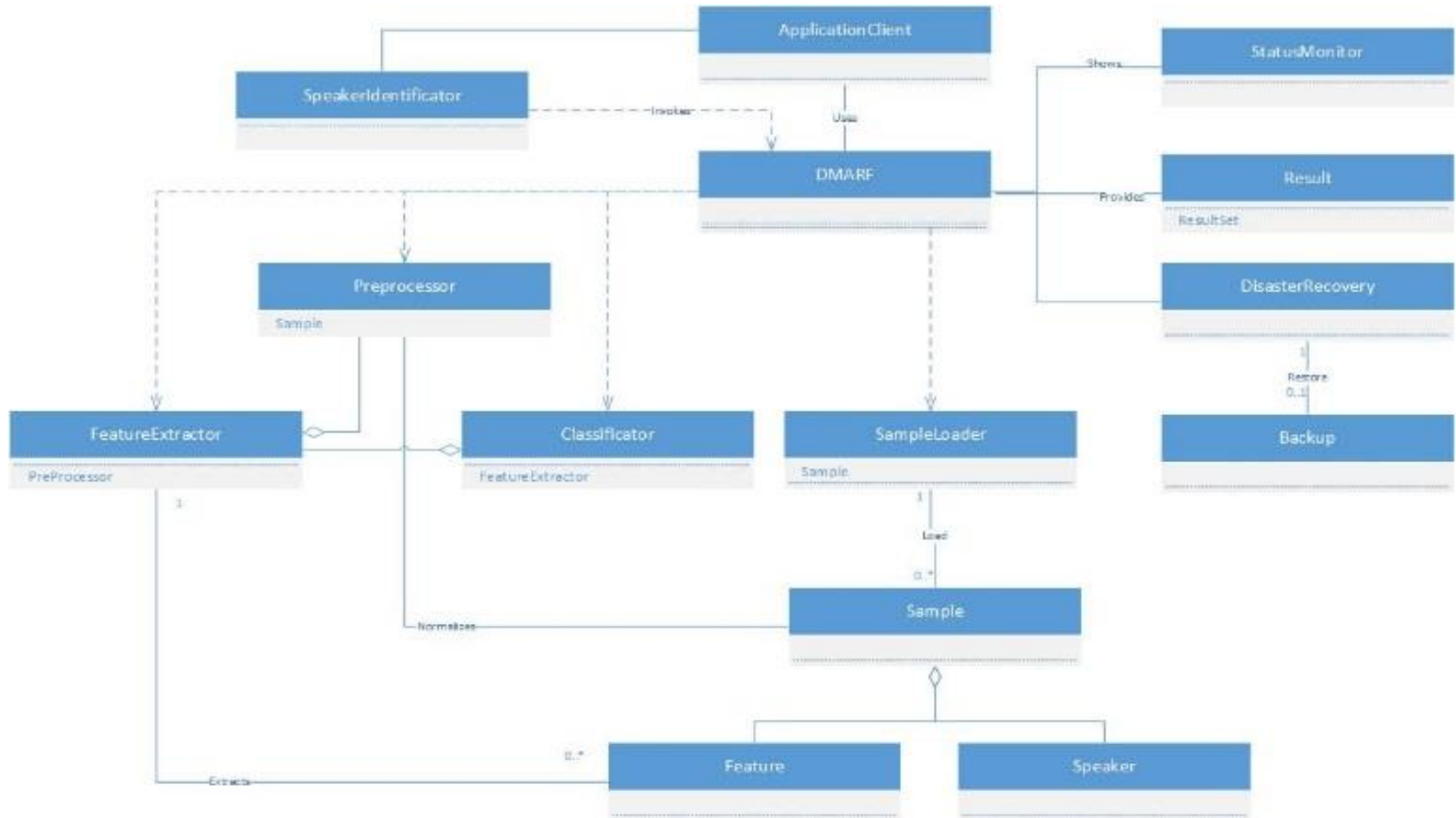

Figure 7 - DMARF Domain Model

**DMARF** includes four stages in its pipeline process as shown in the domain model above. They are represented by the following concept classes: **SampleLoader, Preprocessor, FeatureExtractor**, and **Classificator**.

The process starts by Appliction Client using a SpeakerIdentificator application to invoke DMARF to load one or multiple samples in to SampleLoader. The **Sample** is normalized in the next processing phase by the **Preprocessor**. The **FeatureExtractor** extracts the **Feature** from processed **Sample** using defined methods and algorithms, abstracting the featured vector from **Sample** and making it available to **Classificator**. Classification phase updates the database of training sets with feature vector and implements classification against existing training sets. Finally, the **Result** obtained from each stage is stored using **DisasterRecovery** and any stage can be restored by this service using **Backup**. A **StatusMonitor** service can be implemented to display the status of a **DMARF** service.

### 2) GIPSY

GIPSY mainly contains 4 parts as shown in the domain model below:
1) **RIPE** is the run time environment, which allows the users to communicate with the GIPSY system.
2) **GIPC** is the compiler. The GIPC has individual compilers for each supported language (i.e.: Java, C, CPP, etc.)
3) **GEE** is the execution engine; the demands are generated and are computed in this part.
4) **GEER** is the GEE Resource

**RIPE** contains an Editor and an Inspector responsible for editing inputs/outputs and inspects results before storage respectively.
It is a run-time programming environment that provides the visualization of a dataflow diagram with respect to the Lucid portions of GIPSY programs. RIPE has an interactive nature. Utilizing this nature, the user can draw benefits from RIPE at run-time. It couples with the GIPC and helps user in many ways to allow them in dynamically examining the **Intensional Value Warehouse** (IVW), this module also provides a chance for the user to modify things such as input or output channels of the program.
Communication protocols and components like garbage collector and recompilation of sequential threads can be done through RIPE. The provided Visual diagrams are changed in to a textual form by RIPE and later compiled in to an operational version. This Module identically, is well suited for various kinds of applications.

**GIPC**: This module contains the preprocessor, parsers and compilers which together converts the intensional language and compiles the program. This is considered to be a key module in a GIPSY system because it is built to support Hybrid programming between Intensional Programming Languages (IPL) and Java. This is achieved by introducing the notion of intesional object into the Java language to create a hybrid object-oriented intensional programming language. This module couples with the hybrid language efficiently to determine that the design of the GIPC module helps GIPSY system in achieving goals like flexibility, generality and adaptability. GIPC also enables the "automated generation of frame work hot spots to advance" the overview of the system in supporting the development of programming language.

The **GEE** contains 4 tiers, DGT (Demand Generator Tier) which is responsible for generating new demands (intensional, procedural, system or resource demand), The DWT (Demand Worker Tier) that picks the new demands and performs computation then stores the results in the DST (Demand Storage Tier) which is the storage tier. The GMT (General Management Tier) is the overall system managing the individual tiers. GEE is an execution engine based on a demand driven principle, which means computation takes place only if there is an explicit demand. GEE uses eduction, where computations are implemented in conjunction with a warehouse component. Every demand possibly makes a procedure call that is computed either locally or remotely and also that are concurrent with other demand procedural calls. Every computed value is stored in a warehouse and retrieved back when they are needed; this is because it is cheaper to extract a computed value from a warehouse if needed than computing it once again. GEE also has a demand dispatcher component that makes better decisions in assigning work load to the workers in a balancing way by gathering the required information at run-time [55, 56].

**GEER**: In order to overcome the concept of Language independence of the run- time environment. GIPC depends up on the GEER component that is generated by the compiler. GEER is structured in a generic way that its semantics are of source language independent. GEER also has a dictionary of procedure classes or wrapper classes which wrap the procedures in Java class if necessary [55, 56].

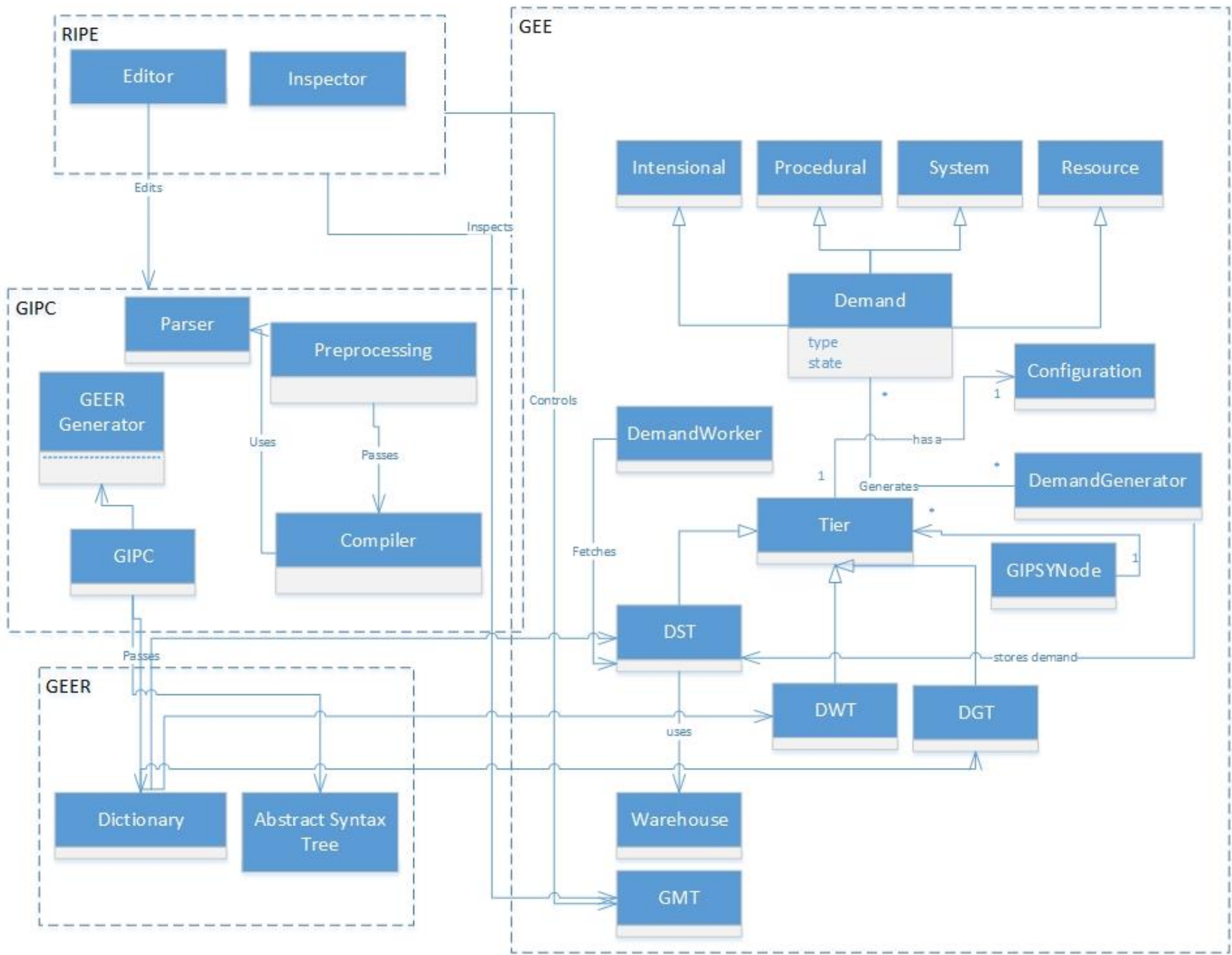

Figure 8 - GIPSY Domain Model

*3) Fused DMARF-over-GIPSY Run-time Architecture (DoGRTA)*

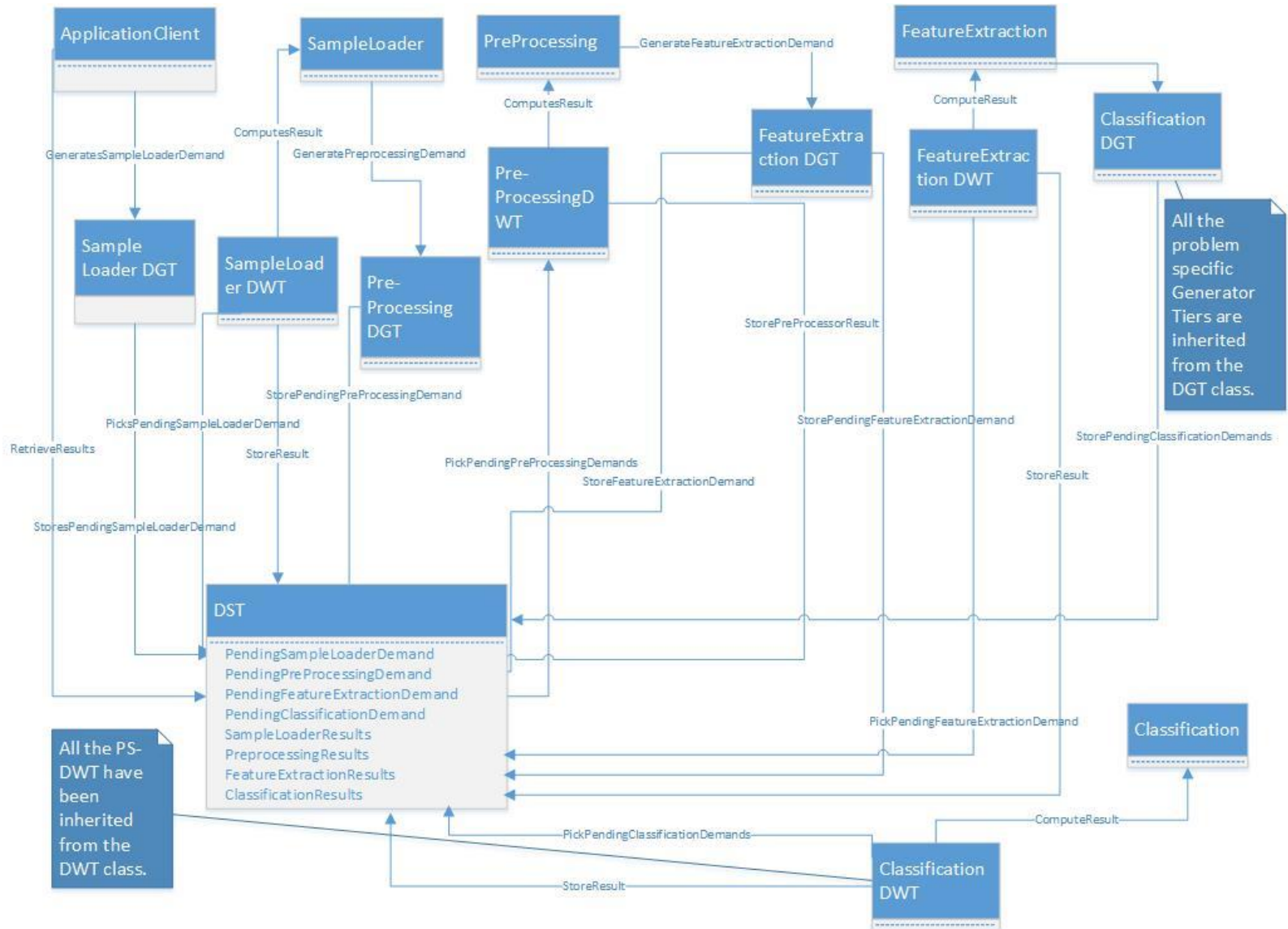

Figure 9 - DMARF-Over-GIPSY Run-time Architecture Domain Model

The generator-worker architecture of GIPSY is integrated to the pipeline of DMARF. The SampleLoader, PreProcessing, FeatureExtraction and Classification of DMARF modules are individually attached with Problem Specific Generator and Worker Tier respectively. All the PS-DGTs are inherited from the DGT class, similarly the PS-DWTs are inherited from the DWT class. The ApplicationClient which provides the input to the DMARF and create a demand for the SampleLoader through the Problem Specific Demand Generator Tier (PS-DGT), this demand will be stored in the storage managed by the DST (Demand storage Tier). The PS-DWT (Problem specific demand worker tier) of SampleLoader picks the pending demands, computes them and stores the result in the DST. Along with the computation, the SampleLoader generates a demand for the PreProcessing through a PS-DGT. The PreProcessing PS-DWT picks the pending demand, performs computation and stores the result back in the DST. Preprocessing will then create a demand for the FeatureExtraction which is computed by the DWT and that result gets stored in the DST. The FeatureExtraction DGT generates the demand for the classification, which will be picked up and computed by DWT. The DST acts as a centralized storage for storing the demands, their status (whether pending or completed) and the computed results. We can also store an additional result independently should a client application be interested in the result of one stage of the DMARF pipeline. The DGT picks the result set, aggregates the information and arranges it in a specific format. These specific results or the complete results can be retrieved by the application client.

*1) DMARF*

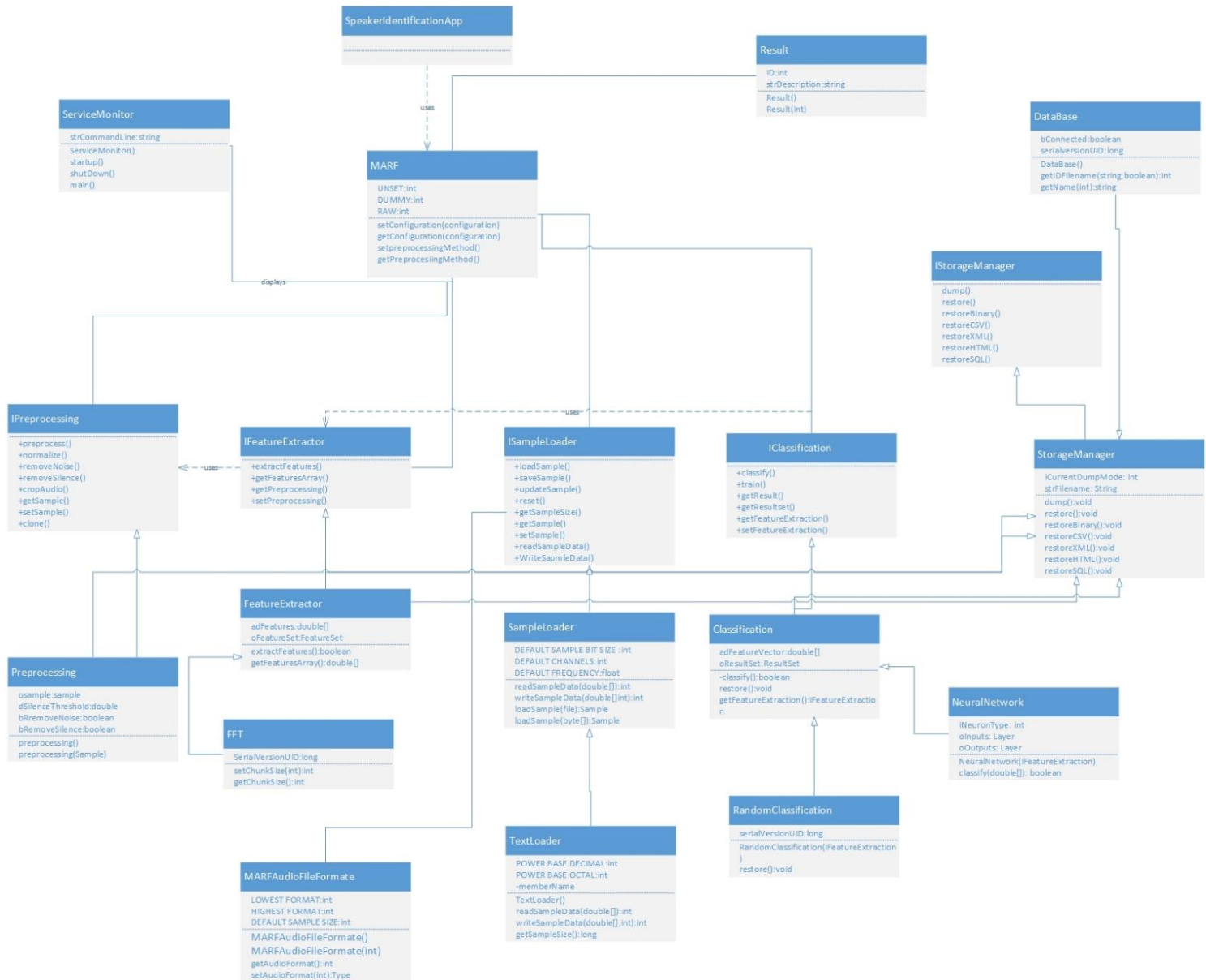

Figure 10 – DMARF Actual Class

The above UML class diagram represents the various important classes presented in DMARF.

DMARF invokes the existing operation of modules for every single state of the pipeline using the it's important interfaces ISampleLoader, IPreprocessing, IFeatureExtractor and IClassification. DMARF contains mainly four sections as Feature Extractor, Preprocessor, SampleLoader and Classification. It starts with the Preprocessor which deals with the input predominantly the audio file. The FeatureExtractor class will extract the features and store it in a Features Array. The SampleLoader interface allows loading of MARFAudioFileFormat which is implemented in the SampleLoader class, while TextLoader implements SampleLoader to modify the sample data to text. Classification class is implemented in classes such as NeuralNetwork, Stochastic, RandomClassification and Distance. ServiceMonitor monitors MARF working. All details, results, storage and backup is handled by the StorageManager. Four classes Preprocessing, FeatureExtractor, Database and Classification implement StorageManager which handle their details. The SpeakerIdenficationApp utilizes the MARF application to identify the speaker based on the analysis of the audio input.

*A. Classes of Interest*

The Modular Audio Recognition Framework (MARF) is an open-source framework built in Java used to recognize and identify audio and classify such identifications by use of pattern recognition algorithms and natural language processing. For the main classes of interest are Application unit, Database, Loader, storage manager, feature extractor. The text loader sends the information to the loader unit. The loader is connected with the application unit which is monitored by the service monitor. The database unit sends information to the storage manager and storage manager is connected to the preprocessing unit and feature extractor. Thus completes the working process of DMARF.

*B. Differences and similarities:*

1.  In the conceptual diagram FeatureExtraction is in aggregation with preprocessing and association with DMARF whereas in actual diagram, FeatureExtraction is in inheritance with StorageManager and association with DMARF.
2.  The Classificator will be in association to DMARF in conceptual diagram but it is just in inheritance with the StorageManager.
3.  The Speaker identifier, SampleLoader by an interface ISampleLoader, preprocessor result set will be in the similar functionality in both the diagrams.

4.  The StatusMonitor in the conceptual diagram is in similar with the actual diagram with the ServiceMonitor.

**Conceptual classes matching to actual classes**

| Conceptual Class | Actual Class |
|---|---|
| DMARF | DMARF |
| Preprocessor | Preprocessor |
| FeatureExtractor | FeatureExtractor |
| SampleLoader | Sample |
| Classificator | Classificator |
| Result | Storage |
| Application Client | Client |
| Speaker identificator | Speakerident |

Table 4 : Conceptual classes mapping to actual classes

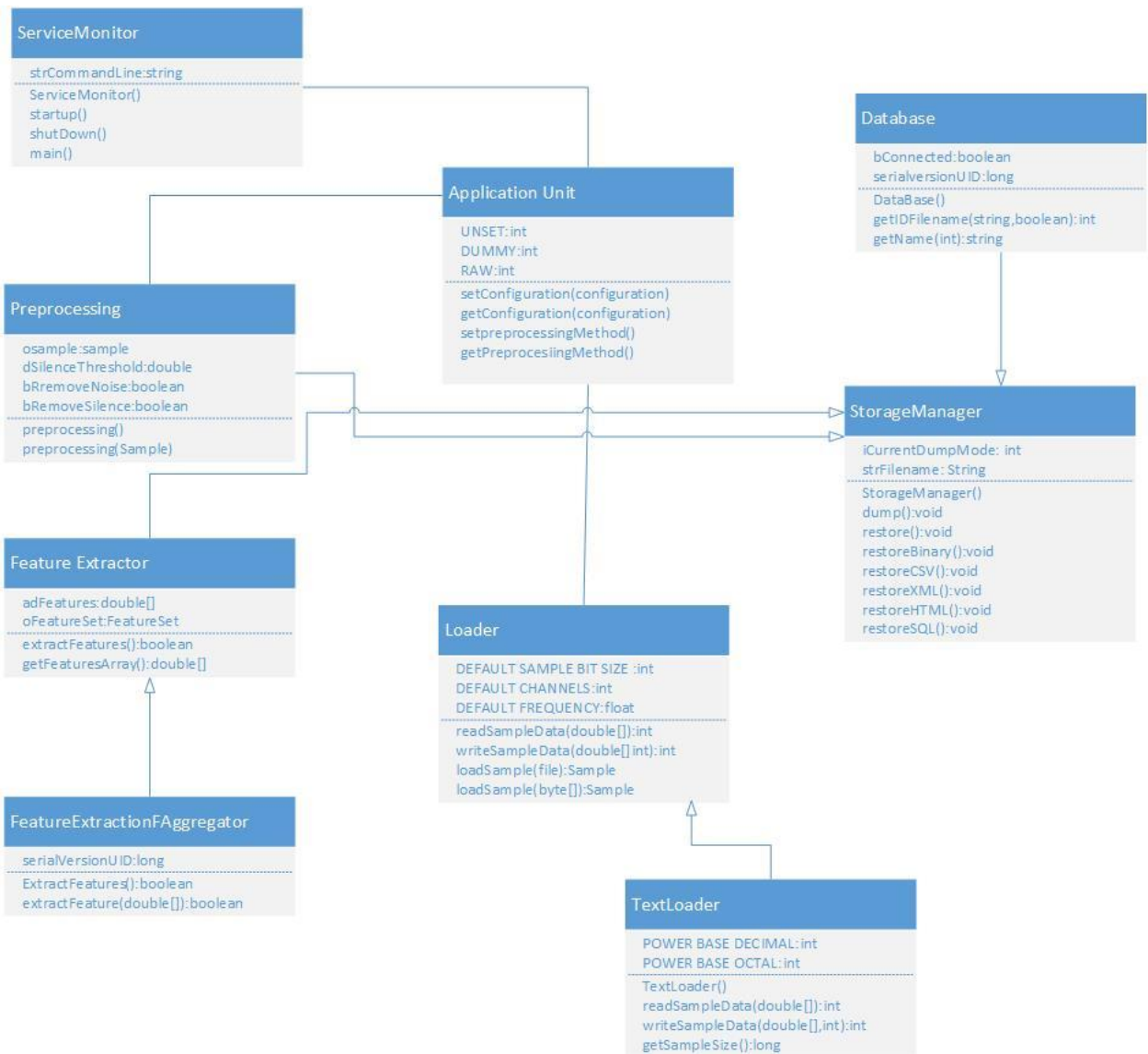

Figure 10: Classes of interest- DMARF

## C. CLASSES AND RELATIONSHIP BETWEEN CLASSES

**Code for Classification.class**

```
package marf.Classification;
import java.util.Vector;
import marf.MARF;
import marf.FeatureExtraction.IFeatureExtraction;
import marf.Storage.ResultSet;
import marf.Storage.StorageException;
import marf.Storage.StorageManager;
import marf.Storage.TrainingSet;
public abstract class Classification
extends StorageManager
implements IClassification
{
        protected IFeatureExtraction oFeatureExtraction =
null;
        protected TrainingSet oTrainingSet = null;
        protected double[] adFeatureVector = null;
        protected ResultSet oResultSet = new ResultSet();
        private static final long serialVersionUID =
7933249658173204609L;
        protected Classification(IFeatureExtraction
poFeatureExtraction)
        {
            this.oFeatureExtraction =
poFeatureExtraction;
                    if(MARF.getModuleParams() != null)
                    {
                        Vector oParams =
MARF.getModuleParams().getClassificationParams();
                        if(oParams != null &&
oParams.size() > 0)
                        {
                                this.iCurrentDumpMode =
((Integer)oParams.elementAt(0)).intValue();
                        }
                    }
        }
```

### Code for RandomClassification.class

```
package marf.Classification.RandomClassification;
import java.util.Random;
import java.util.Vector;
import marf.MARF;
import marf.Classification.Classification;
import marf.Classification.ClassificationException;
import marf.FeatureExtraction.IFeatureExtraction;
import marf.Storage.Result;
import marf.Storage.StorageException;
import marf.util.Debug;
public class RandomClassification
extends Classification
{
    private Vector oIDs = new Vector();
```

```
        private static final long serialVersionUID = -
6770780209979417110L;
        public RandomClassification(IFeatureExtraction
poFeatureExtraction)
        {
                super(poFeatureExtraction);
                this.strFilename = new StringBuffer()

.append(getClass().getName()).append(".")

.append(MARF.getPreprocessingMethod()).append("
.")

.append(MARF.getFeatureExtractionMethod()).appe
nd(".")
                        .append(getDefaultExtension())
                        .toString();
                this.oObjectToSerialize =
this.oIDs;

                }
```

The Classification class is associated with the RandomClassification class directly. The RandomClassification class imports the methods from the Classification class by inheriting it. The Interface IClassification is implemented by the Classification class. As the Classification class imports the methods, the poFeatureExtraction object will be accessed by the methods in the RandomClassification class. The RandomClassification class will train the random sets and they will be stored in the database by the Classification class.

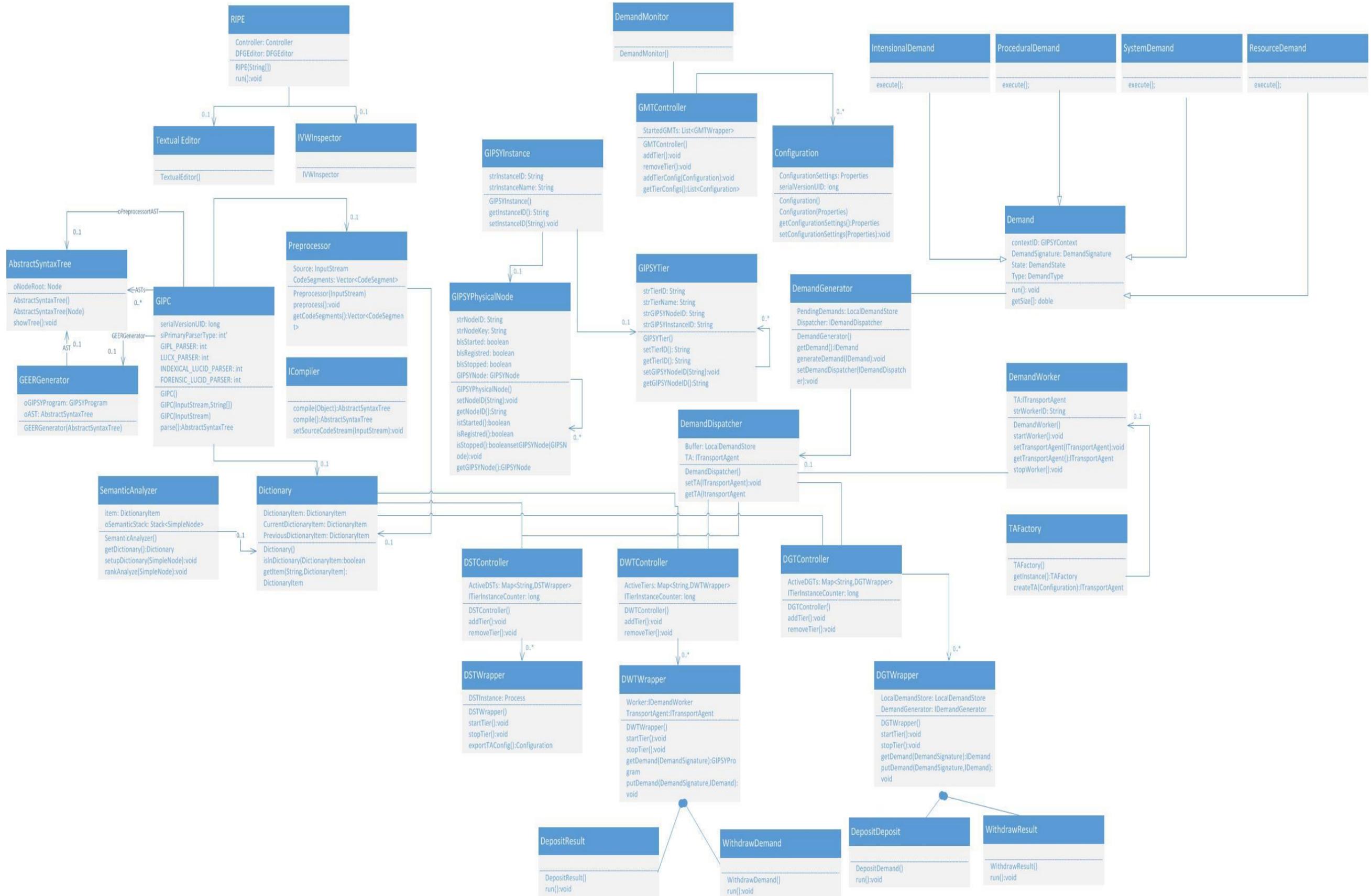

Figure 11 – GIPSY Actual Class Diagram- GIPSY

The class diagram above shows the actual classes of GIPSY. The GIPSY system is sub-divided into 4 main parts

1. GIPC (General Intensional Programming Language Compiler)
2. GEE (General Eduction Engine)
3. RIPE (Run-time Interactive Programming Environment).
4. GEER (General Eduction Engine Resource)

The GIPC class is the compiler class, this class splits the data into LUCID part and sequential part and compiles the LUCID part. The GIPC class is associated with Preprocessor class and Dictionary Class. The preprocessor splits the input program into chunks and feeds to the appropriate LANG parsers, the dictionary class encapsulates the dictionary operations. ICompiler is an interface for the GIPC class in which the actual compilation of

GIPSY code takes place. The compiled part is then passed to the DemandGenerator class of the GEE part. The GEE part has 4 tiers Demand Generator Tier (DGT), Demand Worker Tier (DWT), Demand Storage Tier (DST) and GMT (General Manager Tier). The DemandGenerator is an inner class in DGT class; it creates a unique demand for each computation required. The DemandGenerator class is associated to DemandDispatcher class. The DGTWrapper is a wrapper class to the DGT tier, the class DGTController is assosciated to the DGTWrapper class. The DGTWrapper has 2 subclasses DepositResult and WithdrawDemand. The DemandWorker class is an inner class of DWT tier class; it is responsible for picking up the demands from the storage and performs computations on the demand. The DWTWrapper is the wrapping class; the DWTController is the controlling class which is associated with the DWTWrapper class. The DWTWrapper has 2 subclasses DepositResult and WithdrawDemand. The DSTController class has an association to the DSTWrapper class; these classes are responsible for maintaining the storage of the demands. The GMTController is the overall managing class; it is associated to a configuration class. The RIPE sub-system acts as the visual run time programming environment for the users. The RIPE class allows the user to interact with the GIPSY system. The ripe class has an association relationship with IVWInspector and the TextualEditor classes. The IVWInspector class allows the user to inspect the warehouse and the textual editor allows the user to modify the input/output parameters. The class GIPSYInstance is associated to 2 classes GIPSYPhysicalNode, GIPSYTier.

*A.Classes of Interest*

GIPSY being described as a "demand-driven" architecture, the classes of interest are those dealing with demands which are one of the core attributes of GIPSY. Demands are classified into four types: Intensional, Procedural, Resource and System. All of these classes are inherit from Demand base class.
The DemandType class associated with the Demand class identifies the incoming demand based on type. Such demands are queued up for processing and DemandState class is tasked with assigning the state of the demands. Those waiting to be processed are labelled as 'PENDING' and are queued to the DemandGenerator. Once a demand is taken up by the DemandGenerator its state changes to 'INPROCESS', during which the demand is generated and transferred to the DemandDispatcher. The DemandDispatcher makes use of the TransportAgent to forward the demands to the DemandWorker, where the demands are executed. Executed demands are labelled as 'COMPLETED' and the process starts all over again. There are controllers and monitors that overlook the functioning of this part of the process.

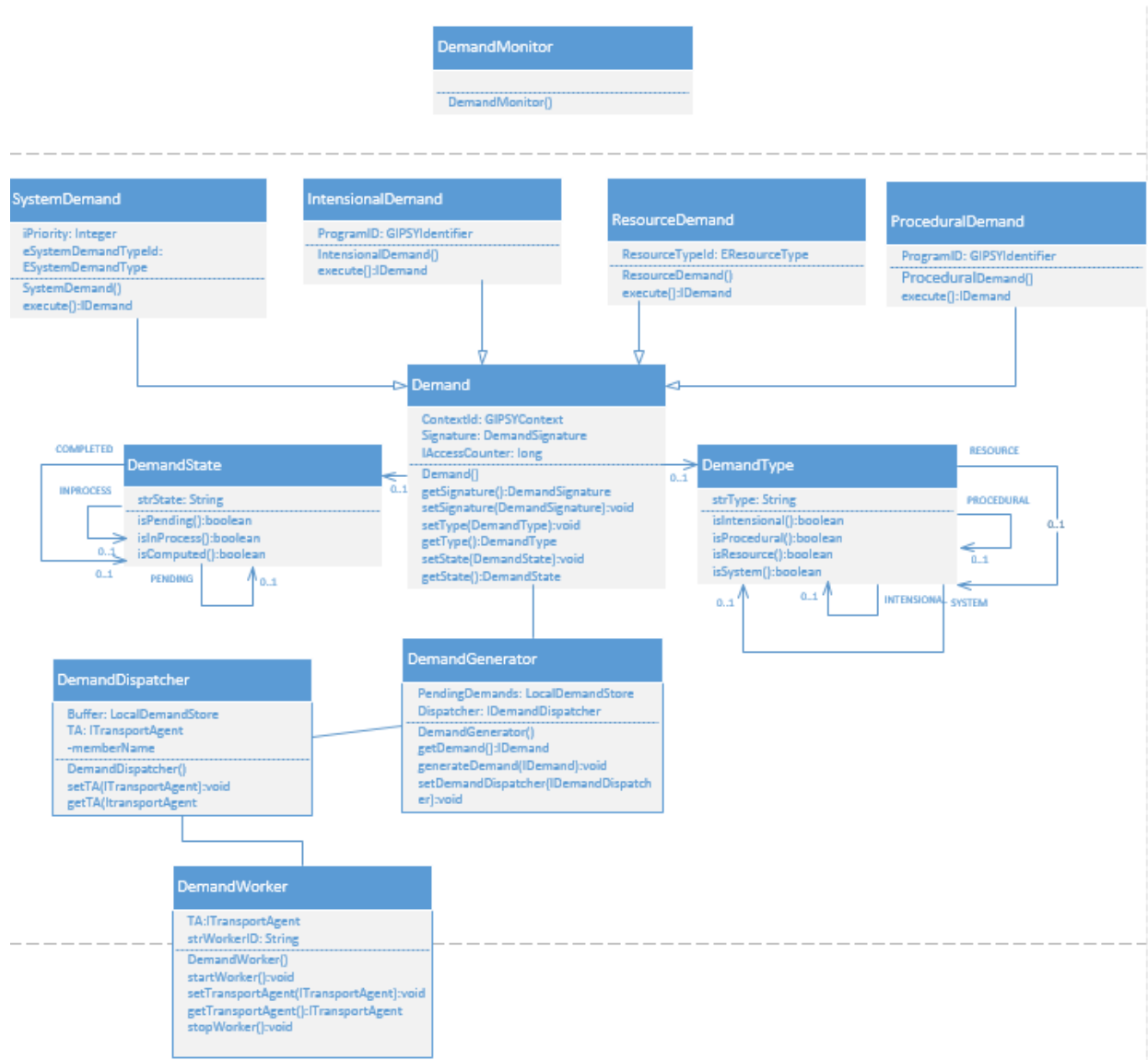

Figure 12 – GIPSY Classes of Interest

- In the Domain diagram the DemandGenerator and DemandWorker accessed the same classes to withdraw demand and store the demand in the storage, whereas in the actual class diagram, for each DemandGenerator class and DemandWorker class separate subclasses DepositResult and WithdrawDemand have been defined.

- In Conceptual classes the high level tiers DGT, DWT, DST and GMT were defined. The actual system the tiers are implemented with 2 classes i) a Controller Class and ii) a Wrapper Class.

- The Parser in the conceptual class is defined using 2 classes i) Dictionary and ii) Preprocessing.

- The conceptual class Warehouse is managed by DST in the actual class diagram.

| Conceptual Class | Actual Class |
|---|---|
| Editor | TextEditor |
| Inspector | IVWInspector |
| Preprocessing | Preprocessor |
| Compiler | Icompiler |
| Configuration | Configuration |
| DemandWorker | DemandWorker |
| DemandGenerator | DemandGenerator |
| DST | DSTController |
| DWT | DWTController |
| DGT | DGTController |
| Warehouse | DSTController |
| GMT | GMTController |
| Tier | GIPSYTier |
| GipsyNode | GIPSYPhysicalNode |

Table 5: Conceptual Classes Mapping to Actual Classes

## C. CLASSES AND RELATIONSHIP BETWEEN CLASSES

**Code for DemandGenerator**

```
package gipsy.GEE.IDP.DemandGenerator;
import
gipsy.GEE.IDP.DemandDispatcher.IDemandDispatcher;

public abstract class DemandGenerator
extends BaseThread
implements IDemandGenerator
{
        protected LocalDemandStore oPendingDemands =
new LocalDemandStore();
        protected IDemandDispatcher oDispatcher = null;
        protected GIPSYProgram oGEER = null;

        public void
setDemandDispatcher(IDemandDispatcher poDispatcher)
        {
                this.oDispatcher = poDispatcher;
        }
}
```

**Code for DemandDispatcher:**

```
package gipsy.GEE.IDP.DemandDispatcher;
import gipsy.GEE.IDP.ITransportAgent;
import gipsy.GEE.multitier.TAExceptionHandler;
import gipsy.interfaces.LocalDemandStore;

public abstract class DemandDispatcher
implements IDemandDispatcher
{

        protected TAExceptionHandler
oTAExceptionHandler = null;
        protected ITransportAgent oTA = null;

        public DemandDispatcher()
        {

        }

        public void
setTAExceptionHandler(TAExceptionHandler
poTAExceptionHandler)
        {
                this.oTAExceptionHandler =
poTAExceptionHandler;
        }
        public TAExceptionHandler
getTAExceptionHandler()
        {
                return this.oTAExceptionHandler;
```

```
        }
        public void setTA(ITransportAgent poTA)
        {
                this.oTA = poTA;
        }
        public ITransportAgent getTA()
        {
                return this.oTA;
        }
}
```

The class DemandGenerator is associated with the DemandDispater class. The interface IDemandDispatcher has been implemented by DemandDispatcher class, and the DemanGenerator class imports the interface IDemandDispatcher to access the class DemandDispatcher. The interface object oDemandDispatcher is defined in the DemandGenerator class to access the DemandDispatcher class. As soon as the demand is generated by the DemandGenerator class, the DemandDispatcher class accesses the demand to either store it in the storage or send it to the DemandWorker.

## VII.   METHODOLOGY

### A. *Refactoring*

#### 1)  *Identification of Code Smells and System Level Refactorings*

Exposure of code smells is the first step to be done during a refactoring process. Code smells are not bugs, but they specify flaws in design that may contribute in breaking down or slowing down development [46]. Detecting code smells is one of the developer's mundane tasks. There are different categories of code smells that are grouped based on their behavior. At the same time there are many tools that aid developers in identifying those code smells.

We discovered different major and minor code smells via the use of tools.  We used the following tools to identify code smells:

**Checkstyle**: Is an Eclipse plugin that helps programmers in a team to write java code that adheres to a coding standard. It helps in detecting Large Class, Long Method and Duplicated Code smells [51].

**JDeodrant:** Is an Eclipse plugin that automatically helps in Identifying Feature Envy, God Class, Long Method and Type Checking code smells in Java programs. The tool even suggests some possible refactoring to the user. [50]

**PMD**: Is an Eclipse plugin that scans java source code for potential problems or possible bugs like dead code, unused local variables and duplicated code [49].

#### a)  *DMARF*

As stated on Metrics, the version of DMARF project analyzed has 125 packages. To find code smells, we could look deeper on these packages and look for more metrics.  We have 1058 classes, where 98.7% are public (1045).

### Code Smells from DMARF are as follows:

1.   **Type Checking:** In this category, code smell can be distinguished into two kinds, class with an attribute that represents state (type field).  The corresponding branch of a conditional statement is executed depending on its value.  If it is a switch statement, the type appears in the switch expression, if it is if or if/else structure, the attributes should be compared for equality with the type field in all conditional expressions.   Classes including (but not limited to) **marf.Storage.ModuleParams, marf.FeatureExtraction.FeatureExtractionFactory, marf.Storage.Loaders.TextLoader** exhibits the Type Checking code smells [52].

2.   **Dead Code:** Portions of code that are not used and will not be deleted, contributes to cost of maintenance of code without producing any benefit is called dead code. DMARF has classes like **ClassificationException.java** that exhibits the property of dead code. This are identified using **CodePro [53].**

3.   **Large class or God Class:** If there is any class that is trying to do too much or having too many responsibilities. These classes have too many instance variables or methods than it is classified as a large class [46]. This is identified by measuring the number of lines of code. Some classes showed signs of God Class code smell as per high number of lines of code. We could identify two as such God classes:

   1.   **marf.MARF (1613 lines of code):**
      - Handling too many responsibilities. Below is a figure that represents this class before refactoring:
      - Using many attributes from external classes.

**a) marf.Classification.NeuralNetwork.NeuralNet work (1114 lines of code):**

- Handling too many responsibilities.

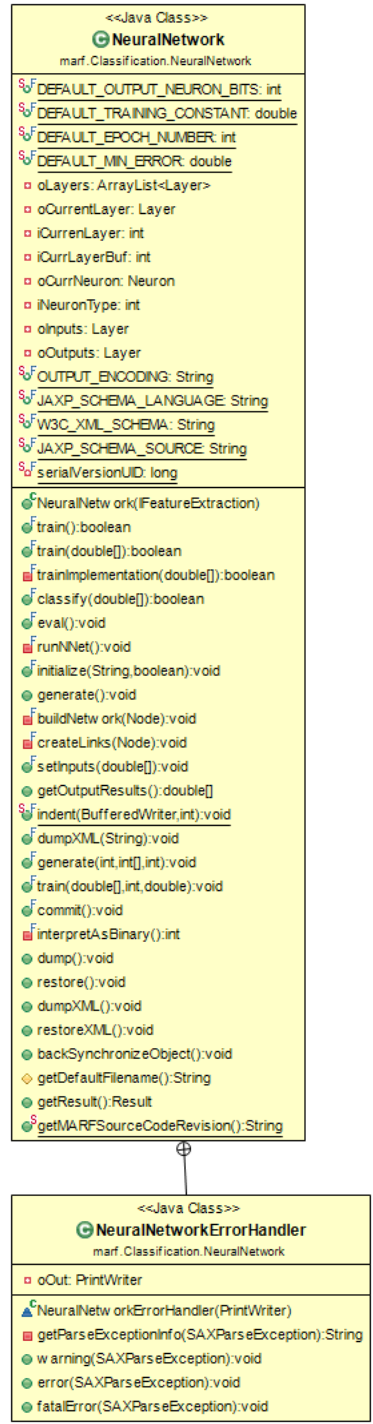

Figure 14: Class NeuralNetwork

**4. Switch Statements**

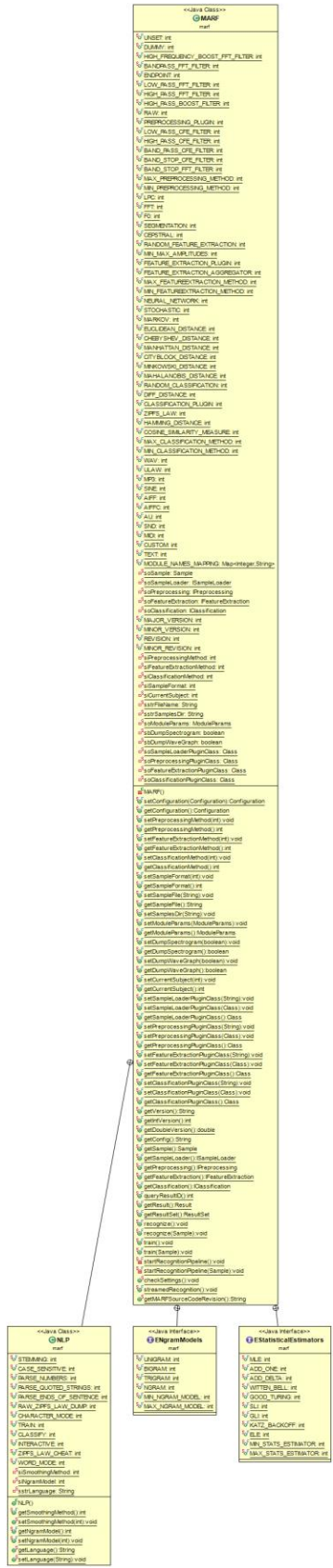

Figure 13 : Class marf.Marf

We found some methods with switch or if/elseif/else statement with many levels of conditional branches. This code smell is a good candidate to apply "Replace Type Code with State/Strategy" refactoring. One such method is the create() method I marf.FeatureExtraction.FeatureExtractionFactory: class. This method has a long switch statement with 10 conditions.

5. **Long Method:** are similar to large classes, but here methods are too long and difficult to understand, modify or extend. Long method tends to centralize the functionality of a class, in the same way as god class does. In the DMARF project, examples are the methods getTrainingSetFilename() in src.marf.Classification.Classification class and create() in src.marf.Classification.ClassificationFactory class. [47].

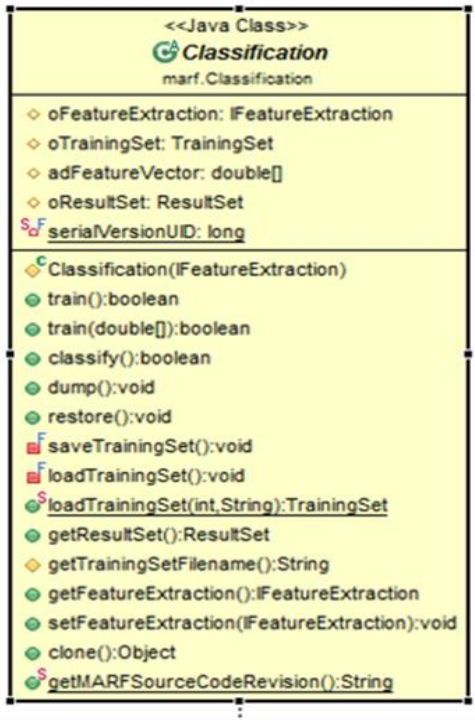

Figure 15: Classification class

*b) GIPSY*

**Code Smells from Gipsy are as follows.**

1. **Feature Envy:** This code smell appears when methods in the class shows more interest in other foreign classes than the one it is placed. This is because of high coupling or tight coupling the methods shows interest in other classes than the one it is located. This is detected by measuring strength of coupling between methods of different classes.

Example:

a. Method named SubTree4 which is originally located in gipsy.GIPC.SemanticAnalyzer class shows more interest in gipsy.GIPC.intensional.SimpleNode class as shown in the code visualization we can say that SubTree4 method indicates all the connections to fields, methods and references in class gipsy.GIPC.intensional.SimpleNode. As shown in Figure 19.

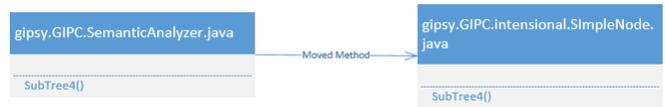

Figure 16: Method SubTree4

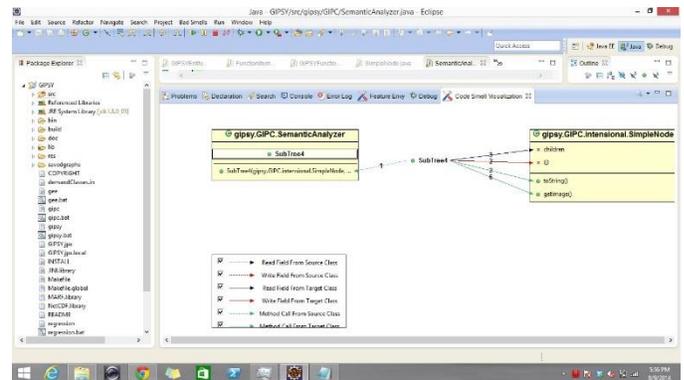

Figure 17 : Code Visualization for Class SemanticAnalyzer

b. **TypeChecking:** Classes like gipsy.lang.GIPSYContext, gipsy.GEE.IDP.DemandWorker.DemandWorker, gipsy.GEE.multitier.GIPSYNode exhibits the Type Checking code smells.

c. **Dead Code:** GIPSY has classes like Interpreter.java, DemandGenerator.java, DemandType.java and many classes that exhibit the property of dead code. This are identified using CodePro [53].

d. **GOD Class:** is also called as design flaw, usually violates the single responsibility principle and it controls a large number of objects implementing different functionalities solution for this is to extract all the methods and fields which are related to specific functionality in to a separate class.

**Example**:

As shown in Figure 20, the visualization shows source class contains all the extract methods and fields. So, in this case **GIPSYNode** class acts as GOD class doing many responsibilities or formed by tight coupling with other code, we need to break the class **GIPSYNode** class in to sub classes.

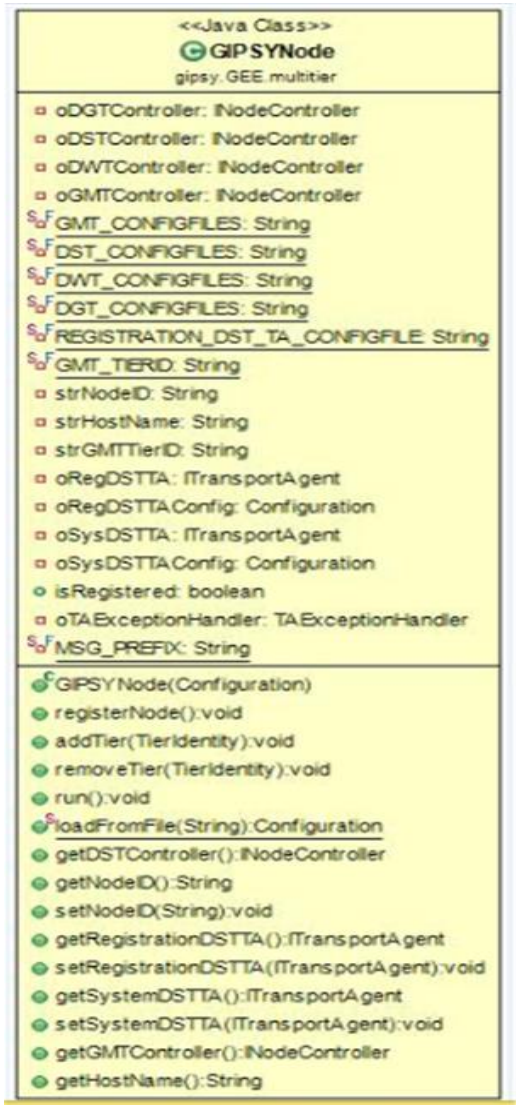

Figure 18: Partial Class Diagram before proposed refactoring

a) We split GIPSYNode class responsibilities to extracted classes.

b) We created two new classes and named it after methods as TierControls and NodeControls

c) We need to link from the GIPSYNode class to the newly formed classes.

d) Moved the constructor code into new method named addTier(), removeTier(),getNode(),setNode()(Extract Method).

e) Created new Classes named TierControls, NodeControls (Extract Class).

f) Modify the code as per proposals.

g) Test the code after the building the new jar file.

h) Note that class NodeControls and TierControls are a candidate for refactoring techniques like Extract method and Extract class but this will not be addressed at this time as our main emphasis on the GIPSYNode class.

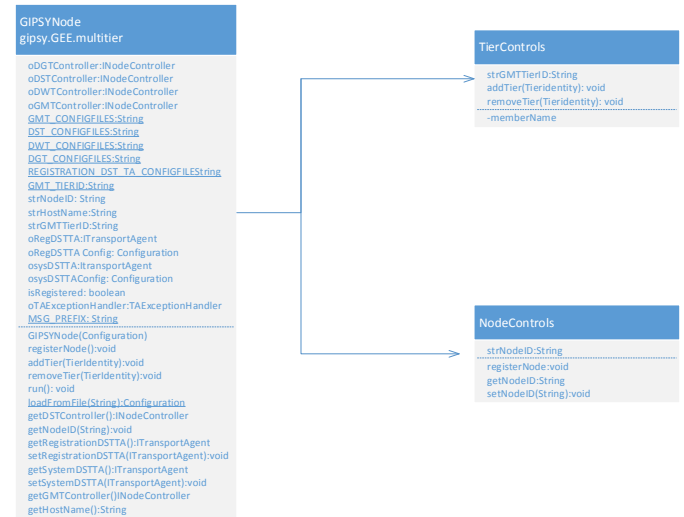

Figure 19: Partial Class Diagram after proposed refactoring

Our prime focus is to increase cohesion and lower coupling of GIPSYNode. We can achieve this by applying refactoring techniques such as Extract Method, Extract Class to GIPSYNode. In applying those techniques, we can move the responsibilities of GIPSYNode to 2 other new classes, introduction of 2 new classes will also increase cohesion.

### 2) *Specific Refactorings will be Implemented in PM4.*

#### a) *DMARF*

We chose 2 classes to apply the refactoring techniques and remove the code smells mentioned in the previous section. The details are followed:

1. FIXING THE LONG METHOD CODE SMELL:

Method getTrainingSetFilename() in class Classification.java (which has 69 lines of code including comments and spaces) has been chosen to remove this particular smell from it. It is not that big in size, but it has a very high level of complexity. It was classified as Cyclomatic Complexity level 11. Cyclomatic Complexity is determined by the number of decision points (as 'if',

'while', 'for' …) in a method plus one for the method entry. Generally, 1-4 is low complexity, 5-7 indicates moderate complexity, 8-10 is high complexity, and 11+ is very high complexity.

To remove these code smells we will apply the following steps:

- Create a new JUnit test case to test Classification class.
- Extract Method: in order to divide the long method into smaller and simpler methods, and give them appropriate names.
- Move Method.
- Compilation and testing is applied after each move.

Once the steps above are completed, we can verify the correctness using the created JUnit test to ensure that it is working properly and the behavior was not changed after refactoring.

### 2. FIXING THE SWITCH STATEMENT CODE SMELL:

For method "setParams()" in class marf.Storage,ModuleParams we will eliminate this complex switch statement Smell by using a HashMap whose index represents which Vector we are setting the parameters. To achieve this we will do the following steps:

- Create a new JUnit test case to test Classification class.
- Declared and initialized a variable HashMap
- Remove the switch statement in the method setParams by adding the parameters to the appropriate Vector indexes by the integer parameter
- Compilation and testing is applied after each move

**Below is the code of the source method:**

```
        private synchronized final void setParams(Vector
poParams, final int piModuleType)
                {
                        if(poParams == null)
                        {
                                throw new
IllegalArgumentException("Parameters vector cannot be
null.");
                        }

                        switch(piModuleType)
                        {
                                case PREPROCESSING:
this.oPreprocessingParams = poParams;
```

```
                                break;

                                case FEATURE_EXTRACTION:

this.oFeatureExtractionParams = poParams;
                                break;

                                case CLASSIFICATION:
this.oClassificationParams = poParams;
                                break;

                                default:
                                        throw new
IllegalArgumentException("Unknown module type: " +
piModuleType + ".");
                        }
                }
```

Once the steps above are completed, we can verify the correctness using the created Junit test to ensure that it is working properly and the behavior was not changed after refactoring.

### b) GIPSY

Based on detected code smells, one possible improvement is the elimination of duplicate switch statements in GIPSYNode class under the gipsy.GEE.multitier package. Both methods run a similar conditional check to determine to which controller to either add or remove a tier from. The goal is to extract this implementation into its own method so that it is only present in one part leading to better maintainability and evolution in the case that we add a new type of tier and it's respective controller.. The code dump below illustrates this implementation for the addTier (TierIdentity peTierIdentity) and removeTier (TierIdentity peTierIdentity) methods:

**Below is the code of the source method:**

```
public void addTier(TierIdentity peTierIdentity)
{
        switch (peTierIdentity)
        {
                case DGT:
                {
                        this.oDGTController.addTier();
                break;
                }
                case DST:
                {
                        this.oDSTController.addTier();
                break;
                }
                case DWT:
                {
                        this.oDWTController.addTier();
```

```
            break;
          }
      }

}

public void removeTier(TierIdentity peTierIdentity)
{
      switch (peTierIdentity)
      {
            case DGT:
            {

            this.oDGTController.removeTier();
                  break;
            }
            case DST:
            {

            this.oDSTController.removeTier();
                  break;
            }
            case DWT:
            {

            this.oDWTController.removeTier();
                  break;
            }
      }
}
```

Once the implementation is complete, we will verify the correctness by running JUnit test cases to ensure that we are still able to add and remove the correct tiers and thus the behavior has not changed.

## B. Identification of Design Patterns

A pattern detection application byNikolaos Tsantalis was used called pattern4 [43]. The reverse engineering tool used to visualize the classes and their relationships in the pattern identification process is ObjectAid UML [42]. Below is a table summarizing which team member identified which pattern in which project.

| Design Patterns | | |
|---|---|---|
| **Team** | **DMARF** | **GIPSY** |
| 1) Pavan Kumar Polu (p_polu) | | Observer Pattern |
| 2) Gustavo Pereira (gu_perei) | Adapter Pattern | |
| 3) Amjad Al Najjar (a_alna) | | Singleton Pattern |
| 4) Prince Japhlet (p_stephe) | | Chain Of Responsibility pattern |
| 5) Biswajit Banik (bi_banik) | Factory pattern | |
| 6) Bhanu Prakash R. (b_ramine) | State-strategy pattern | |
| 7) Ajay Sujit Kumar | | Decorator pattern |
| 8) sabari Krishna Raparla | Composite pattern | |

Table 6 : Design Patterns Contribution

### 1)  DMARF

#### a)  Adapter Design Pattern

One of the most used design pattern in the MARF project is the Adapter pattern. The Adapter converts the interface of a class into another interface clients expect. Adapter lets classes work together that couldn't otherwise because of incompatible interfaces. Brief, it wraps an existing class with a new interface [44]. It is used to make existing classes work with others without modifying their source code.

There are two types of Adapter Pattern: Object Adapter Pattern and Class Adapter Pattern. In Object Adapter Pattern, the adapter contains an instance of the class it wraps. So, the adapter makes calls to the instance of the wrapped object.

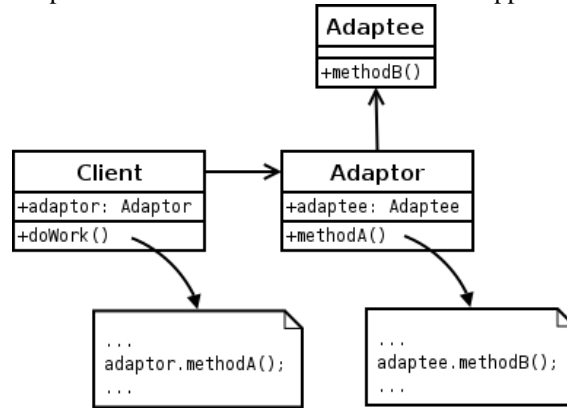

Figure 20: Object Adapter Pattern [45]

The Class Adapter Pattern the adapter is created by implementing or inheriting both the interface that is expected and the interface that is pre-existing, using multiple polymorphic interfaces [45].

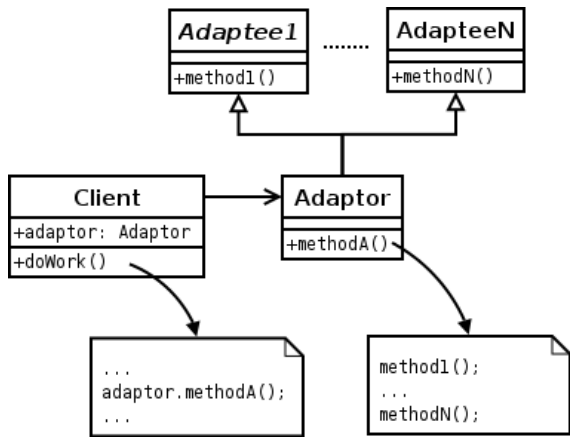

Figure 21: Object Adapter Pattern [45]

One example of Object Adapter Pattern in DMARF is represented by the Class Diagram below:

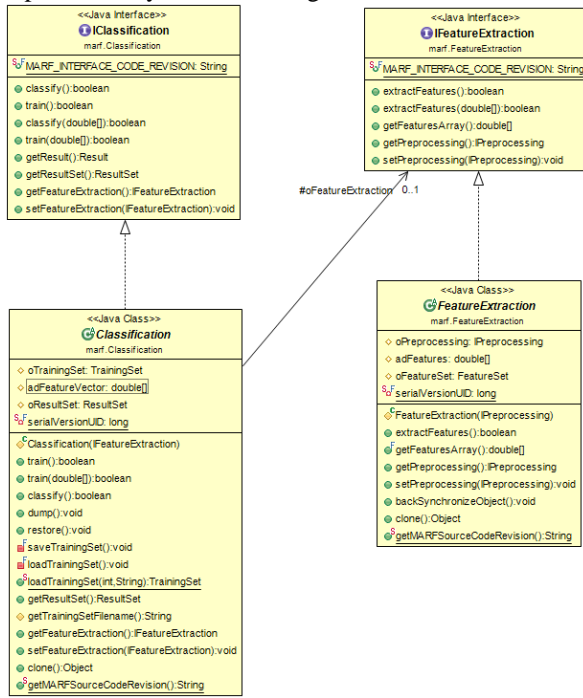

Figure 22 : UML Class Diagram for Adapter Pattern

The interface IFeatureExtraction is the Adaptee/Receiver of the Adapter class Classification. The adapter hides the adaptee's interface from the client. The goal here is to provide an interface for each class that inherits class attributes.

**Below is the class iClassification source code and its attributes:**

```
package marf.Classification;
import marf.FeatureExtraction.IFeatureExtraction;
import marf.Storage.Result;
import marf.Storage.ResultSet;

public interface IClassification
{

        String MARF_INTERFACE_CODE_REVISION =
"$Revision: 1.1 $";
        boolean classify()
        throws ClassificationException;
        boolean train()
        throws ClassificationException;
        boolean classify(double[] padFeatureVector)
        throws ClassificationException;
        boolean train(double[] padFeatureVector)
        throws ClassificationException;
        Result getResult();
        ResultSet getResultSet();
        IFeatureExtraction getFeatureExtraction();
        void        setFeatureExtraction(IFeatureExtraction
poFeatureExtraction);
}
```

The methods that Request ( )/Execute ( ) from the Adapter/ConcreteCommand class Classification are train ( ): boolean and classify ( ): boolean.

*b) Factory Design pattern*

**Factory pattern:** The main goal of the factory pattern is to define an interface for creating objects. It instantiate new objects without using new directly it prefers until runtime to decide the kind of objects that needs to be instantiated [58]. It generally creates a pure fabrication of factory that mainly handles the object creation. The two main intents of factory pattern are creating objects without exposing the instantiation logic to the client and referring to the newly created object through a common interface [57].

In order to create new object the implementation process works like the following way when client needs a product, but instead of creating it directly using the new operator, it asks the factory object for a new product, providing the information about the type of object it needs.

In DMARF case study, the factory pattern occurs in the project is here. The class FeatureExtractionFactory implements the factory pattern.

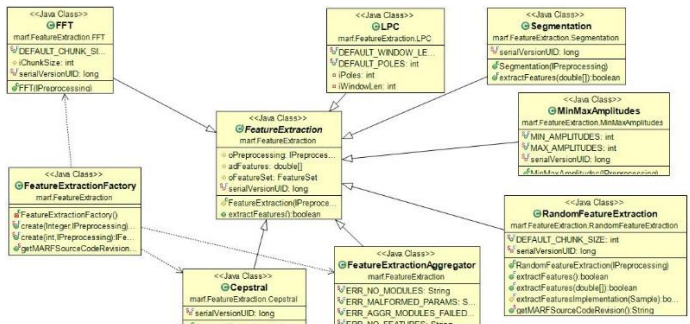

**Figure 23: Factory pattern**

Feature Extraction is acting as a factory for the classes Segmentation, LPCm Ceptral, Aggregator, MinMaxAmplitudes, FFT, and RamdomFeatureExtarction.

**Below is the relevant code dump of the FeatureExtractionFactory class:**

```
public static final IFeatureExtraction create(final int
piFeatureExtractionMethod, IPreprocessing poPreprocessing)
throws FeatureExtractionException
{
IFeatureExtraction oFeatureExtraction = null;

switch(piFeatureExtractionMethod)
{
case MARF.LPC:
oFeatureExtraction = new LPC(poPreprocessing);
break;

case MARF.FFT:
oFeatureExtraction = new FFT(poPreprocessing);
break;

case MARF.F0:
oFeatureExtraction = new F0(poPreprocessing);
break;

case MARF.SEGMENTATION:
oFeatureExtraction = new Segmentation(poPreprocessing);
break;

case MARF.CEPSTRAL:
oFeatureExtraction = new Cepstral(poPreprocessing);
break;

case MARF.RANDOM_FEATURE_EXTRACTION:
oFeatureExtraction = new
RandomFeatureExtraction(poPreprocessing);
break;

case MARF.MIN_MAX_AMPLITUDES:
oFeatureExtraction = new
MinMaxAmplitudes(poPreprocessing);
```

```
break;

case MARF.FEATURE_EXTRACTION_PLUGIN:
{
try
{
oFeatureExtraction =
(IFeatureExtraction)MARF.getFeatureExtractionPluginClass()
.newInstance();
oFeatureExtraction.setPreprocessing(poPreprocessing);
}
catch(Exception e)
{
throw new FeatureExtractionException(e.getMessage(), e);
}
break;
}
case MARF.FEATURE_EXTRACTION_AGGREGATOR:
{
oFeatureExtraction = new
FeatureExtractionAggregator(poPreprocessing);
break;
}
default:
{
throw new FeatureExtractionException
(
"Unknown feature extraction method: " +
piFeatureExtractionMethod
);
}
}
return oFeatureExtraction;
}
public static String getMARFSourceCodeRevision()
{
return "$Revision: 1.3 $";
}
}
```

*c) Composite Design pattern*

Composite Design Pattern represents the whole – part hierarchies in to tree structures to compose objects. This pattern allows users to treat objects and their composition uniformly, composite pattern structured in a way that has four elements component, leaf, composite, and client. Component which is an interface and acts as an abstraction for leafs and composites [35, 36]. Component defines that interface must be implemented by the objects in composite. A composite stores child components in addition to apply methods that are defined by the component interface. Leafs are object with no children and they provide services explained by the Component interface. Client is the one who operates the objects using component interface. We generally use this pattern whenever we have "composites that contains components, each of which could be composite" [35]. A user has a situation to a tree data structure and desires to complete functionality operations on all nodes independent of the fact that a node might be a branch or a leaf. The user basically attains reference to the essential node by using the component interface, and deals with the node using this interface; it doesn't bother if the node is a leaf or composite.

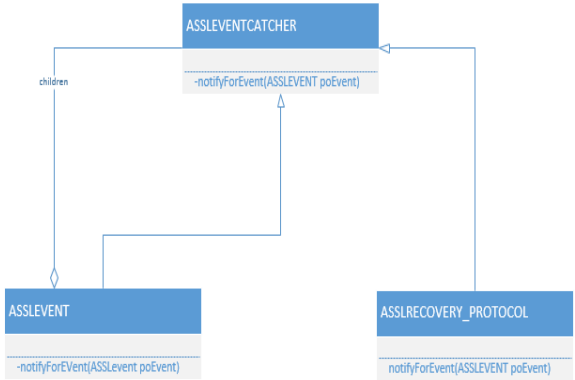

Figure 24: UML Class Diagram for Composite Pattern

**COMPONENT:**
```
public interface ASSLEVENTCATCHER
{
   public void notifyForEvent ( ASSLEVENT poEvent );
}
```

COMPOSITE:
```
public class ASSLEVENT extends Thread implements
ASSLEVENTCATCHER, ASSLMESSAGECATCHER
{
    public synchronized void notifyForEvent ( ASSLEVENT
poEvent )
        {
                vOccurredEvents.add(poEvent);
        }
}
```

**LEAF:**
```
public class ASSLRECOVERY_PROTOCOL  implements
ASSLEVENTCATCHER
{
    public synchronized void notifyForEvent ( ASSLEVENT
poEvent )
        {
            Enumeration<ASSLEVENT>     eEVENTS     =
vInitiatedByEvents.elements();
                ASSLEVENT currEvent = null;
                while ( eEVENTS.hasMoreElements() )
```

```
                {
                        currEvent                    =
eEVENTS.nextElement();
                        if ( currEvent == poEvent )
                        {
                                save();
                                break;
                        }
                }
        }
}
```

*d) State-strategy Design pattern*

The state pattern explains the behavior of an object depends on the change of its state. For an object defined in one class the behavior will be changing for that particular object on the change in its other classes [54]. The strategy pattern will be of the giving a solution for the varying but related classes or algorithms by connecting them with a common interface. In DMARF, the classes FeatureExtraction and Preprocessing are connected with an interface IPreprocessing. The object poprocessing is given as an internal reference in the FeatureExtraction class for the Preprocessing class. The behavior of removeNoise and removeSilence are changing in the Ipreprocessing and in Preprocessing classesclasses. Thus the State-Strategy pattern is identified in the following classes

**Code for FeatureExtraction**:
```
package marf.FeatureExtraction;
import marf.Preprocessing.IPreprocessing;
public abstract class FeatureExtraction
extends StorageManager
implements IFeatureExtraction
{
        protected IPreprocessing oPreprocessing = null;
        protected FeatureExtraction(IPreprocessing
poPreprocessing)
        {
                this.oPreprocessing = poPreprocessing;
                this.iCurrentDumpMode =
DUMP_GZIP_BINARY;
                this.oObjectToSerialize = this.adFeatures;
        }
}
```

**Code for IPreprocessing**:
```
package marf.Preprocessing;

import marf.Storage.Sample;

public interface IPreprocessing
extends Cloneable
{
```

String MARF_INTERFACE_CODE_REVISION =
"$Revision: 1.7 $";

```
        boolean preprocess()
        throws PreprocessingException;
        boolean removeNoise()
        throws PreprocessingException;
        boolean removeSilence()
        throws PreprocessingException;
        boolean normalize()
        throws PreprocessingException;
        throws CloneNotSupportedException;
}
```

**Code for Preprocessing:**
```
package marf.Preprocessing;
public abstract class Preprocessing
extends StorageManager
implements IPreprocessing
{
        public final static double
DEFAULT_SILENCE_THRESHOLD = 0.001;
        protected Preprocessing(IPreprocessing
poPreprocessing)
        throws PreprocessingException
        {
                if(poPreprocessing == null)
                {
                        throw new
IllegalArgumentException("Preprocessing parameter cannot
be null.");
                }
        }
}
```

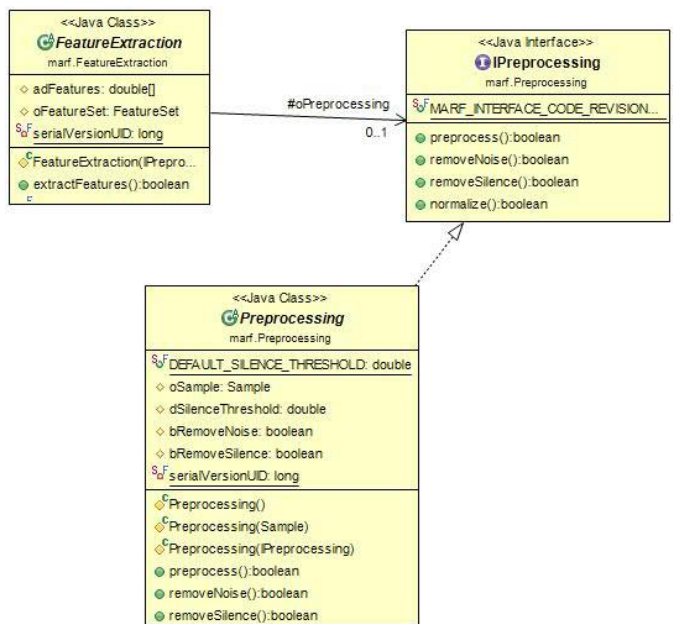

Figure 25: Classes used to implement State Strategy Pattern

## 2)  GIPSY

### a)  Singleton Design Pattern

There are many occurrences of the Singleton design pattern in the GIPSY project. One example is the TAFactory class in the gipsy.GEE.multitier.DST package. As described by Craig Larman, the goal of a singleton is to provide exactly one instance of a class accessible from a single point [Larman 348, 349]. The class itself defines a method *getInstance()* which would return an instance of that class. The class's constructor is private so that instance cannot be created outside of it. In this scenario, the goal was to have only one instance of the TAFactory based on configurations properties and save instantiation effort. Thus any class requiring the TAFactory would get the same instance along with the configurations previously set. It is important to note that the TAFactory instance is only initialized when the *getInstance()* method is invoked the first time. This practice is referred to as lazy initializing and is preferred since initialization is only performed if the class is actually needed therefore saving resources that may be expensive [Larman 350]. Since GIPSY is a multi-threaded project, the creation step of the TAClass is **synchronized** to ensure only one instance is ever created. Figure 18 below shows the relationship between the TAFactory and any class that invokes it's *getInstance()* method.

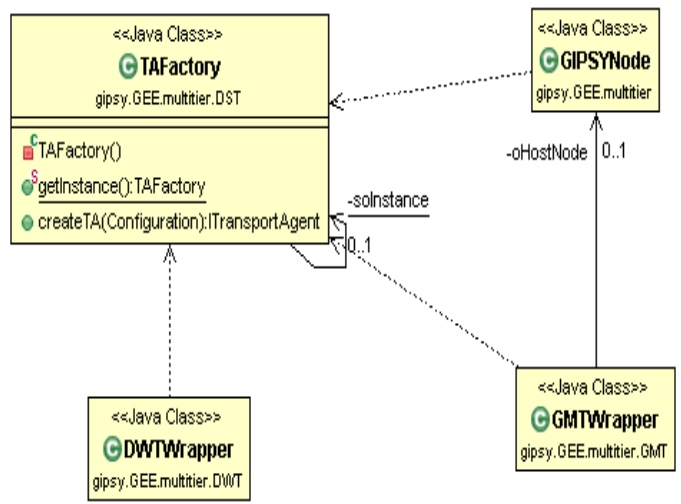

Figure 26 : UML Class Diagram for Singleton Pattern

To identify a simple pattern like Singleton, it was sufficient to search the project for "getInstance" method somewhat unique to this pattern and examine the methods to draw a candidate.

Below is the class TAFactory and any important declarations and methods.

public class TAFactory

```
{
        private static TAFactory soInstance = null;

        private TAFactory()
        {

        }

        public synchronized static TAFactory getInstance()
        {
                if(soInstance == null)
                {
                        soInstance = new TAFactory();
                }
                return soInstance;
        }
}
```

### b) Observer Design Pattern

There are two instance of the observer design pattern in the GIPSY project. Observer pattern defines a one-to-many dependency where the subject class notifies changes on the object by calling one or more methods in the observer class. [37]

Whenever there is a state change in the object of the subject class JJTForensicLucidParserState, the notify method closeNodeScope(Node n, boolean condition) notifies the Observer class Node by accessing the Observer's methods jjtSetParent and jjtAddChild. The Observer class being an interface, these methods are in turn implemented by its child class SimpleNode.

**Example:-**

We can identify the Observer pattern in the following package

**gipsy.GIPC.util**

**Implementation**

**Observer**: gipsy.GIPC.util.Node

**Subject:**
gipsy.GIPC.intensional.SIPL.ForensicLucid.JJTForensicLucid ParserState

**Notify():** gipsy.GIPC.intensional.SIPL.ForensicLucid.JJTFore nsicLucidParserState::closeNodeScope(gipsy.GIPC.util.Node, int):void

**Notify():** gipsy.GIPC.intensional.SIPL.ForensicLucid.JJTFore nsicLucidParserState::closeNodeScope(gipsy.GIPC.util.Node, boolean):void

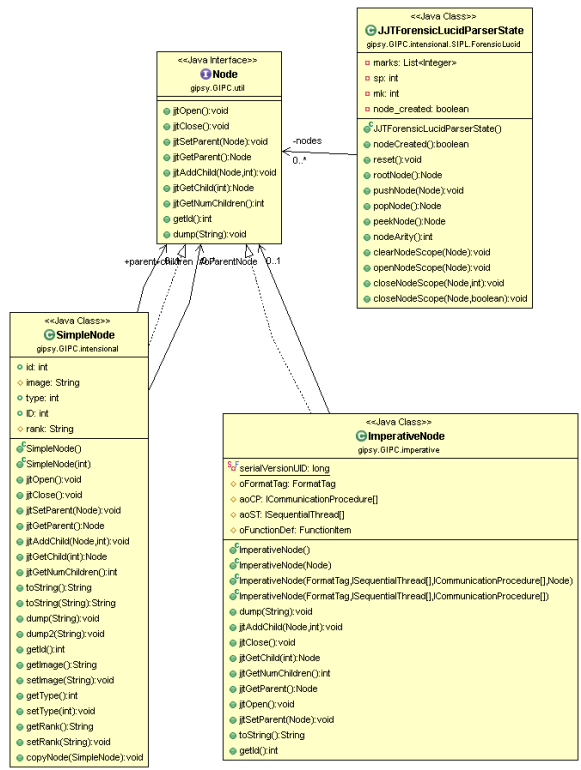

Figure 27: UML Class Diagram for Observer Pattern

Below is the class JJTForensicLucidParserState and any important declarations and methods.

```
public class JJTForensicLucidParserState {

public void closeNodeScope(Node n, boolean condition) {
  if (condition) {
    int a = nodeArity();
    mk = marks.remove(marks.size()-1);
    while (a-- > 0) {
      Node c = popNode();
      c.jjtSetParent(n);
      n.jjtAddChild(c, a);
    }
    n.jjtClose();
    pushNode(n);
    node_created = true;
  } else {
    mk = marks.remove(marks.size()-1);
    node_created = false;
  }
}

}
```

public interface Node
extends Serializable

```
{

    public void jjtSetParent(Node n);

    public void jjtAddChild(Node n, int i);
}

public class SimpleNode implements Node
{
  public void jjtSetParent(Node n) { parent = n; }
  public Node jjtGetParent() { return parent
  public void jjtAddChild(Node n, int i) {
    if (children == null) {
      children = new Node[i + 1];
    } else if (i >= children.length) {
      Node c[] = new Node[i + 1];
      System.arraycopy(children, 0, c, 0, children.length);
      children = c;
    }
    children[i] = n;

  }
}
```

    *c) Decorator Design Pattern*

The class DemandWorker in GIPSY has a decorator pattern. The decorator design pattern comes under structural patterns as it acts as a wrapper for the existing class [39, 40]. The decorator pattern allows addition of new functionality to an already existing object without having to change the structure [40] me and hence there will not be change in the structure [39, 40]. Unlike inheritance where a change in any class would affect the other classes, using decorator pattern any single object of a class can be selected and its behavior can be modified leaving the other instances as they are [41].

The DemandWorker implements the classe IDemandWorker which creates the blue print of the DemandWorker class, this interface has the declarations of the run-time functionalities implemented within the DemandWorker class [Code 1]. The decorator class MARFPCATDWT acts as a wrapper class and as the actual implementations of the decorator functions [Code 2]. The decorator class has 4 run-time operations i) setTransportAgent ii) StartWorker iii) StopWorker and iv) setTAExceptionHandler. The decorator pattern is required to implement the above operations at run-time without changing the structure of the DemandWorker class.

**Code for interface:**
```
public interface IDemandWorker
extends Runnable
{
        void        setTransportAgent(EDMFImplementation
poDMFImp);
        void setTransportAgent(ITransportAgent poTA);
        void    setTAExceptionHandler(TAExceptionHandler
poTAExceptionHandler);
```

```
        void startWorker();
        void stopWorker();
}
```

Code for DemandWorker:

```
        Public void
setTransportAgent(EDMFImplementation poDMFImp)
        {

        this.oDemandWorker.setTransportAgent(poDMFImp
);
        }

        public void setTransportAgent(ITransportAgent
poTA)
        {

        this.oDemandWorker.setTransportAgent(poTA);
        }
        public void startWorker()
        {
                this.oDemandWorker.startWorker();
                this.bIsWorking = true;

        }
        public void stopWorker()
        {
                this.oDemandWorker.stopWorker();
                this.bIsWorking = false;

        }
        public void
setTAExceptionHandler(TAExceptionHandler
poTAExceptionHandler)
        {

        this.oDemandWorker.setTAExceptionHandler(poTA
ExceptionHandler);
        }
```

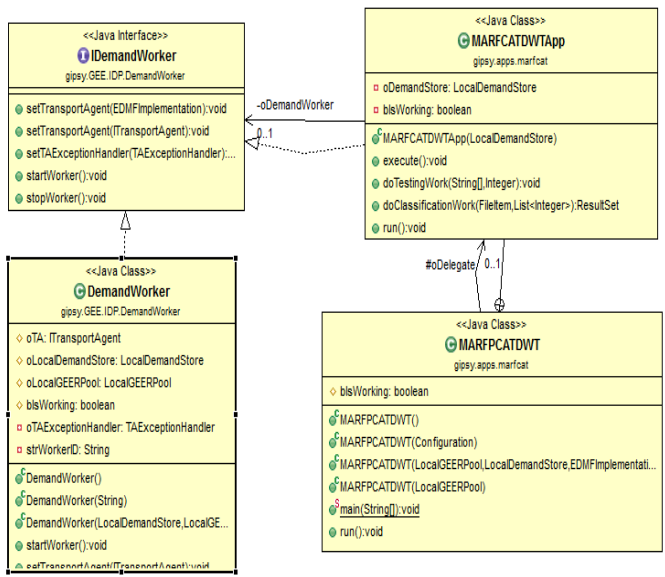

Figure 28: UML Class Diagram for Decorator Pattern [42]

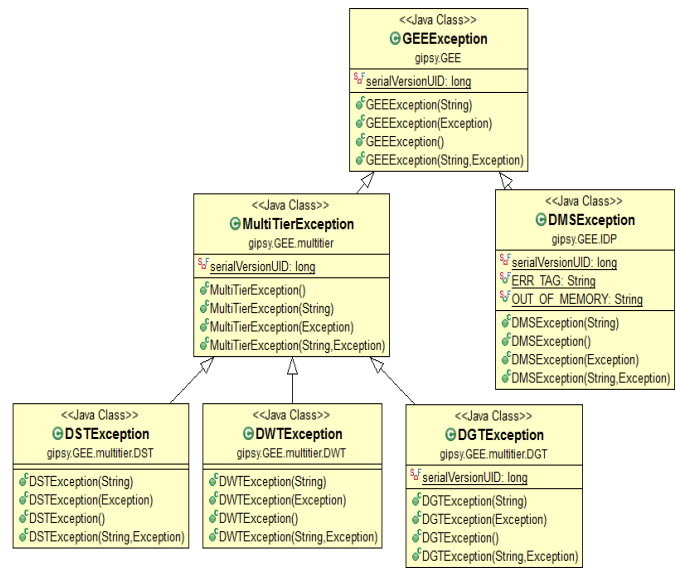

Figure 29: UML Class Diagram for Chain of Responsibility

*d) Chain of Responsibility Pattern*

The Chain of Responsibility pattern contains a series of objects that could potentially handle the request sent to the receiver. This enhances the decoupling of the request sender to the receiver by allowing opportunity for multiple objects along the chain to handle the request in question. [1] It is not mandatory upon receiver to know the structure of the chain or which object will handle the request. The request is passed until an object takes responsibility for the request and overrides appropriate methods where necessary. This pattern is used where it is acceptable to have a receiver object not willing to take responsibility and passes the responsibility to the next object. When none of the objects are willing to take responsibility, a generic handler can be used to negate the possibility of the request being left unhandled.

In GIPSY case study, chain of responsibility pattern occurs multiple times in the project. It is hard to miss the occurrence of the pattern in the exception handling section of the GIPSY Java code [2]. There are a series of classes that extends in a tree-form from the base exception class. Each of the classes are capable of handling specific exception else pass the responsibility to the next higher level exception handler class in the tree.

The GEEException class in turn is an extension of GIPSYException, which extends MARFException finally routing to the Java library class Exception. Similar to GEE Exception handling GIPC and RIPE exception handling use Chain of Responsibility pattern.

**Code for GEEException:**
public class GEEException
extends GIPSYException
{

    private static final long *serialVersionUID* = 8364742916487908785L;

    public GEEException(String pstrMessage)
    {
        super(pstrMessage);
    }

    public GEEException(Exception poException)
    {
        super(poException.getMessage(),
poException);
    }

    public GEEException()
    {
        super();
    }

    public GEEException(String pstrMessage, Exception poException)
    {
        super(pstrMessage, poException);
    }
}

public class MultiTierException
extends GEEException
{

```
        private static final long serialVersionUID = -
6195320385652857639L;

        public MultiTierException()
        {
                super();
        }

        public MultiTierException(String pstrMessage)
        {
                super(pstrMessage);
        }
        public MultiTierException(Exception poException)
        {
                super(poException);
        }

        public    MultiTierException(String    pstrMessage,
Exception poException)
        {
                super(pstrMessage, poException);
        }
}
```

## VIII. IMPLEMENTATION

### A. Refactoring Changesets and Diffs

*1) DMARF*
*Refactoring 1:*

**Change 0/2 – Refactoring Refactor Simplify Conditional Expressions on Method setParams on ModuleParams class**
We eliminate the switch statement CodeSmell by using a HashMap whose index represents which Vector we are setting the parameters. Before the refactoring and after, we are running JUnit tests that are written in first step.

**Change 1/2 – Write JUnit tests for ModuleParams class**
The first step is to write the JUnit tests so that we can validate our changes. There were no specific JUnit tests cases for ModuleParams class, so we wrote the following tests specific to the method that we intended to refactor.

```
public class ModuleParamsTest extends TestCase {

        @Test
        public void testSetParams() {
                final ModuleParams moduleParams = new
ModuleParams();
                //Setup preprocessing vector
preProcessingVectors.put(FEATURE_EXTRACTION,
oFeatureExtractionParams);

        preProcessingVectors.put(CLASSIFICATION,
oClassificationParams);
```

```
        final Vector preProcessingParams = new
Vector();

        preProcessingParams.add("1");
                //Setup feature extraction vector
                final Vector featureExtractionParams = new
Vector();

        featureExtractionParams.add("2");

                //Setup classification Vector
                final Vector classificationParams = new
Vector();

        classificationParams.add("3");
                //Bind appropriate vector to the module
params
                moduleParams.setPreprocessingParams(preProcessin
gParams);
                moduleParams.setFeatureExtractionParams(featureE
xtractionParams);
                moduleParams.setClassificationParams(classification
Params);
                //Verify correct elements were added to the
right vectors
                assertEquals("1",
moduleParams.getPreprocessingParams().get(0));
                assertEquals("2",
moduleParams.getFeatureExtractionParams().get(0));
                assertEquals("3",
moduleParams.getClassificationParams().get(0));
        }
}
```

Our tests are actually indirectly testing the **private** method setParams. Each of those methods invokes setParams from within moduleParams class. Since we did not want to change the visibility of setParams, we determined that testing the method that invokes them is sufficient.

**Change 2/2 – Refactor setParams method in ModuleParams class**
The goal was to eliminate the use of switch statement by using a HashMap whose index represents which Vector we are setting the parameters for (Preprocessing, Feature_Extraction or Classification Vector). To achieve this, we declared and initialized a HashMap as follows:

```
private HashMap<Integer, Vector> preProcessingVectors =
new HashMap<Integer, Vector>();

        {
        preProcessingVectors.put(PREPROCESSING,
oPreprocessingParams);

        }
```

Next we eliminated the switch statement in setParams by simply adding the parameters to the appropriate Vector indexes by the integer parameter piModuleType. The figure below shows this update.

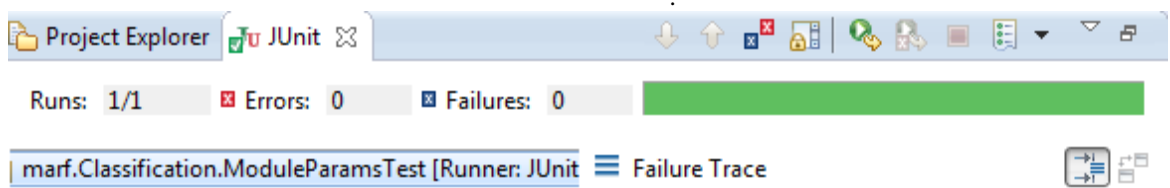

**Figure 30: Code diff for ModuleParams class**

Finally, we reran the tests we wrote in step 1 to verify that our changes are working.

**Figure 31: Test after refactoring**

*Refactoring 2:*

### Change 0/5: Refactor Simplify Conditional Expressions on Method *getTrainingSetFilename(   )* from class *src/marf/Classification/Classification.java*

In Classification class, we created a new method *ProcessData(  )* to be able to reduce the Cyclomatic Complexity from method *getTrainingSetFilename*. We move the switch statement to the new method. We also needed to pull up the the variables commom for both methods, that are iNoiseRemoved and iSilenceRemoved.

### Change 1/5: Create and run ClassificationTest class

Create a JUnit TestCase for Classification class and run it showing it is working properly. It will also be used to ensure that behaviour will not be changed during refactoring.

```
public class ClassificationTest extends TestCase {

    @Test
    public void testGetTrainingSetFilename_1()
        throws FeatureExtractionException {
```

```
LPC lpc = new LPC(new Dummy());
Classification fixure = new DiffDistance(lpc);
lpc.extractFeatures(new double[] { 1, 2, 3, 4, 5 });

String result = fixure.getTrainingSetFilename();
assertNotNull(result);
}
}
```

Run JUnit test before changes:

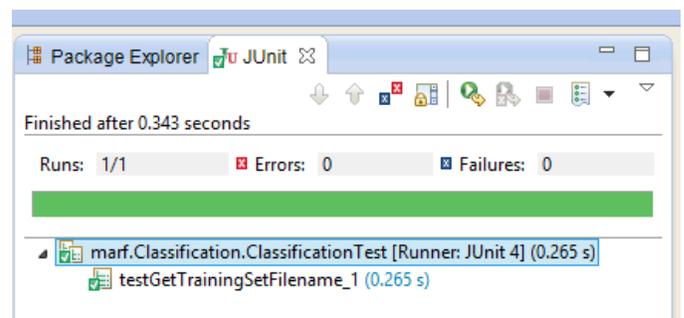

**Figure 32: Test Execution before changes**

## Change 2/5: Create new method *ProcessData( )*

Create a new method *ProcessData( )* in class *src/marf/Classification/Classification.java* to be able to move code from *getTrainingSetFilename( )*

```
+        private void processData(Vector oPreprocessingParams) {
+ }
```

## Change 3/5: Extract switch statement logic to the new method

Extract the switch statement logic from getTrainingSetFilename( ) and add in the new method new method ProcessData( ). Remove it from old method:

```
        private void processData(Vector oPreprocessingParams) {

                switch(oPreprocessingParams.size())
                {
                case 0:
                        break;
                case 1:
                        {
                                if(oPreprocessingParams.firstElement() instanceof Boolean)
                                {
                                        boolean temp = ((Boolean)oPreprocessingParams.firstElement()).booleanValue();
                                        iNoiseRemoved = temp == true ? 1 : 0;
                                }
                                break;
                        }
                default:
                        {
                                if(oPreprocessingParams.firstElement() instanceof Boolean)
                                {
                                        iNoiseRemoved = ((Boolean)oPreprocessingParams.firstElement()).booleanValue() == true ? 1 : 0;
                                }
                                if(oPreprocessingParams.elementAt(1) instanceof Boolean)
                                {
                                        iSilenceRemoved = ((Boolean)oPreprocessingParams.elementAt(1)).booleanValue() == true ? 1 : 0;
                                }
                                break;
                        }
                }
        }
```

```
        if(oPreprocessingParams != null)
        {
                switch(oPreprocessingParams.size())
                {
                case 0:
                        break;
                case 1:
                        {
                                if(oPreprocessingParams.firstElement() instanceof Boolean)
                                {
                                        iNoiseRemoved = ((Boolean)oPreprocessingParams.firstElement()).booleanValue() == true ? 1 : 0;
                                }
                                break;
                        }
                default:
                        {
                                if(oPreprocessingParams.firstElement() instanceof Boolean)
                                {
                                        iNoiseRemoved = ((Boolean)oPreprocessingParams.firstElement()).booleanValue() == true ? 1 : 0;
                                }
                                if(oPreprocessingParams.elementAt(1) instanceof Boolean)
                                {
                                        iSilenceRemoved = ((Boolean)oPreprocessingParams.elementAt(1)).booleanValue() == true ? 1 : 0;
                                }
                                break;
                        }
                }
```

Here is the diff of getTrainingSetFilename method:

```
public String getTrainingSetFilename()
{
    // Prevents NullPointerException, bug #1539695
    int iFeaturesCount = this.oFeatureExtraction == null
        ? (this.adFeatureVector == null ? 0 : this.adFeatureVector.length)
        : this.oFeatureExtraction.getFeaturesArray().length;

    if(MARF.getModuleParams() != null)
    {
        Vector oPreprocessingParams = MARF.getModuleParams().getPreprocessingParams();

        if(oPreprocessingParams != null)
        {
            processData(oPreprocessingParams); // Refactoring DPARF 1
        }
    }
```

```
int iSilenceRemoved = 0;

if(MARF.getModuleParams() != null)
{
    Vector oPreprocessingParams = MARF.getModuleParams().getPreprocessingParams();

    if(oPreprocessingParams != null)
    {
        switch(oPreprocessingParams.size())
        {
            case 0:
                break;

            case 1:

                if(oPreprocessingParams.firstElement() instanceof Boolean)
                {
                    iNoiseRemoved = ((Boolean)oPreprocessingParams.firstElement()).booleanValue()
                }

                break;

            default:

                if(oPreprocessingParams.firstElement() instanceof Boolean)
                {
                    iNoiseRemoved = ((Boolean)oPreprocessingParams.firstElement()).booleanValue()
                }

                if(oPreprocessingParams.elementAt(1) instanceof Boolean)
                {
                    iSilenceRemoved = ((Boolean)oPreprocessingParams.elementAt(1)).booleanValue()
                }

                break;
        }
    }
}
```

**Figure 33: Code diff for method getTrainingSetFilename**

## Change 4/5: Pull up common variables

Pull up the commom variables for both methods, that are iNoiseRemoved and iSilenceRemoved to the Classification class.

```
=public abstract class Classification
=extends StorageManager
=implements IClassification
={
+      /* Data Members */
+      // For comparison, distinguish samples with or without

+      // noise and silence removed
+      int iNoiseRemoved = 0;
+      int iSilenceRemoved = 0;

=      /**
```

ClassificationTestfor to ensure that behaviour was not changed during refactoring.

```
=      * Reference to the enclosed FeatureExtraction object.
=      */

=            :
this.oFeatureExtraction.getFeaturesArray().length;

            // For comparison, distinguish samples with or without

            // noise and silence removed
            int iNoiseRemoved = 0;
            int iSilenceRemoved = 0;

=            if(MARF.getModuleParams() != null)
```

## Change 5/5: Run ClassificationTest class

Run                          JUnit                          TestCase

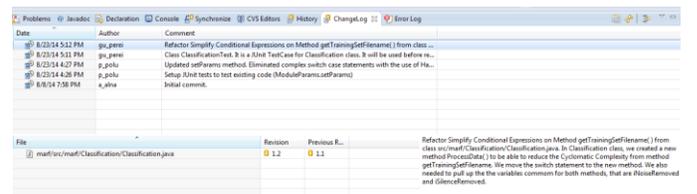

**Figure 34: ChangeLog**

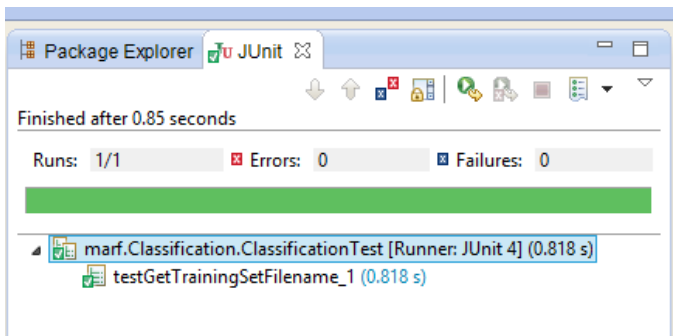

**Figure 35: Test after refactoring**

*1)  GIPSY*

**Change 0/3 – Extract Method refactoring**

We will Refactor duplicate code Code Smell with Extract Method refactoring. We will build up a TestCase and run it before and after all steps.

**Change 1/3 – Write JUnit tests for GIPSYNode class**

Before beginning to refactor, we verified the completeness of the existing JUnit tests. Unfortunately, it only contained the method skeleton and nothing more. It appears that the JUnit tests to GIPSYNode were never implemented. So the first step was to write some tests to assert the current behavior of addTier and removeTier methods. Next, we ran the tests and made sure they passed successfully. The figure below shows the base and our update of the JUnit test cases:

**Figure 36:  Code diff for testAddTier**

**Change 2/3 – Refactor duplicate code into its own method**
In this step, we began with the refactoring process. We can observe that in both addTier() and removeTier() we have a common switch statement that is checking the TierIdentity to determine which INodeController should either add or remove a tier. As this duplicated behavior is common, the idea is to extract this into its own method to centralize this logic into one location making potential future changes easy. To achieve this, we created a getControllerbyTierIdentity taking for parameter the TierIdentity. We moved the common switch statement within this new method and returned the appropriate concrete class that implemented the INodeController interface. Below is the code for the described method:
public INodeController
getControllerByTierIdentity(TierIdentity peTierIdentity){

    switch (peTierIdentity)
    {
        case DGT:
        {

            return oDGTController;
        }
        case DST:
        {
            return oDSTController;
        }
        case DWT:
        {
            return oDWTController;
        }
    }

**Change 3/3 – Update addTier and removeTier to use new method**
Next, we updated the addTier and removeTier to call this new method passing the TierIdentity parameter to it and assigning the result to the INodeInterface object. Then, each can proceed to either add or remove tier as normal. The figure below illustrates the changes to these two methods:

**GIPSYNode.java 1.2**

```java
    public void addTier(TierIdentity peTierIdentity)
    {
        INodeController oController = getControllerByTierIdent
        oController.addTier();
    }
```

**GIPSYNode.java 1.1**

```java
    public void addTier(TierIdentity peTierIdentity)
    {
        switch (peTierIdentity)
        {
            case DGT:
            {
                this.oDGTController.addTier();
                break;
            }
            case DST:
            {
                this.oDSTController.addTier();
                break;
            }
            case DWT:
            {
                this.oDWTController.addTier();
                break;
            }
        }
    }
```

**Figure 38: Code diff of the addTier class**

**GIPSYNode.java 1.2**

```java
    public void removeTier(TierIdentity peTierIdentity)
    {
        INodeController oController = getControllerByTierIdent
        oController.removeTier();
    }
```

**GIPSYNode.java 1.1**

```java
    public void removeTier(TierIdentity peTierIdentity)
    {
        switch (peTierIdentity)
        {
            case DGT:
            {
                this.oDGTController.removeTier();
                break;
            }
            case DST:
            {
                this.oDSTController.removeTier();
                break;
            }
            case DWT:
            {
                this.oDWTController.removeTier();
                break;
            }
        }
    }
```

**Figure 39: Code diff of the removeTier class**

Finally, we reran the tests we had written in step 1 and verified that they still pass successfully.

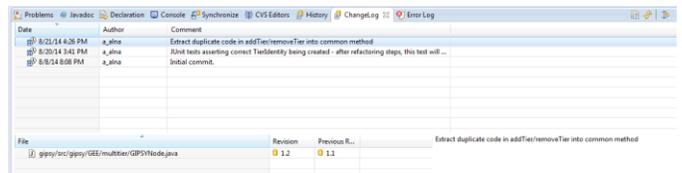

**Figure 42: ChangeLog**

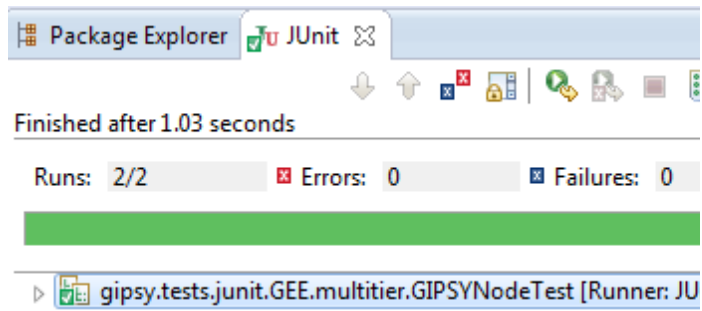

**Figure 40: Test cases**

## IX. CONCLUSION

This research project helped the team members in the analysis and understanding of the architecture design principles and patterns using two excellent case studies for this purpose: DMARF and GIPSY. The overall architecture became clearer throughout the project development and after deep analysis of classes and structures on last stages of the project. The background theory on architectures and design patterns was essential to the team be able to find code smells and apply refactoring techniques. It was a great opportunity to develop programming skills on Java and get knowledge on using repository and versioning control system as CVS. Eclipse was used as IDE and we had the opportunity to experiment with many related tools (plugins) such as CodePro, Analytix, SonarQube, InCode ObjectAid UML, JDeodorant and PMD. All practical work was great to refresh our knowledge of design patterns and refactoring techniques.

## X. ACRONYMS

| MARF | Modular Audio Recognition Framework |
|---|---|
| DMARF | Distributed Modular Audio Recognition Framework |
| ADMARF | Autonomic Distributed Modular Audio Recognition Framework |
| GIPSY | A General Intensional Programming System |
| AGIPSY | Autonomous GIPSY |
| UML | Unified Modeling language |
| PMD | Programming Mistake Detector |
| NLP | natural language processing |
| API | Application Program interface |
| RMI | Remote Method Invocation. |
| ASSL | Autonomic System Specification Language |
| AS | Autonomic Systems |
| AE | Autonomic Elements |
| AC | Autonomic Computing |
| SLO | Service Level Objectives |
| RPC | Remote Procedure Call |
| HTTP | HyperText Transfer Protocol |
| TCP | Transmission Control Protocol |
| XML-RPC | eXtensible Markup Language-Remote Procedure Call |

| SNMP | Simple network management protocol |
|---|---|
| HOIL | Higher-Order Intensional Logic |
| MIB | Management Information Base |
| SMI | Structure of Management Information |
| ANS | Abstract Syntax Notation |
| JDSF | Java Data Security Framework |
| SFAP | Self-forensics autonomic property |
| JOOIP | Java-based object oriented intensional programming language |
| RIPE | Run-time Interactive Programming Environment |
| GIPC | General Intensional Programming Language Compiler |
| GEE | Generic Eduction Engine |
| GEER | Generic Eduction Engine Resources |
| GN | Gipsy Nodes |
| ICP | Intensional Communication Procedures |
| IDS | Intensional Data Dependency Structure |
| PS | Problem Specific |
| DGT | Demand Generator Tier |
| DWT | Demand Worker Tier |
| DST | Demand Store Tier |
| GMT | General Manager Tier |
| IVW | Intensional Value Warehouse |
| IPL | Intensional Programming Language |
| DMF | Demand Migration Framework |
| TA | Transport Agents |
| JMS | Java Messaging Service |
| C | Programming language |
| GUI | Graphical User Interface |
| CVS | Concurrent Versioning System |
| IDE | Integrated development environment |

APPENDIX

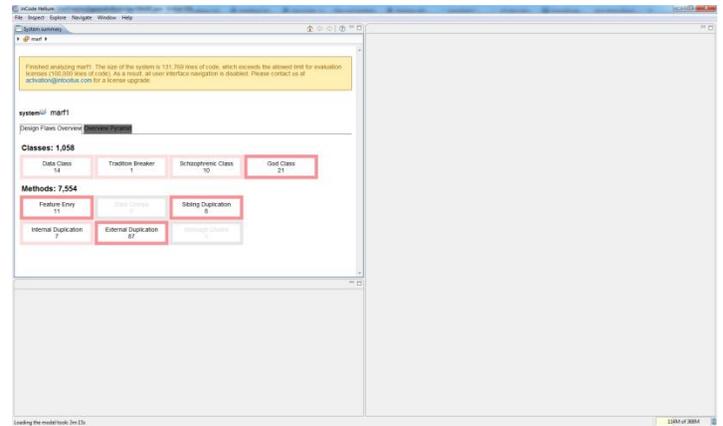

Figure 41: InCode estimates for DMARF

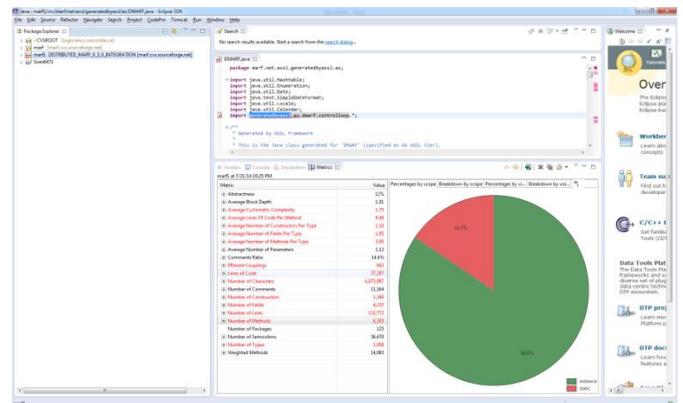

Figure 42: CodePro estimates for DMARF

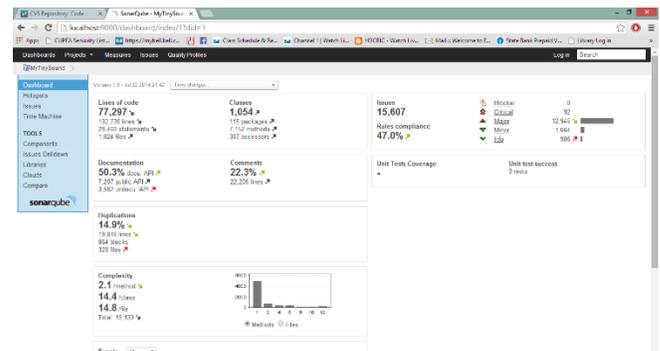

Figure 43: SonarQube estimates for DMARF

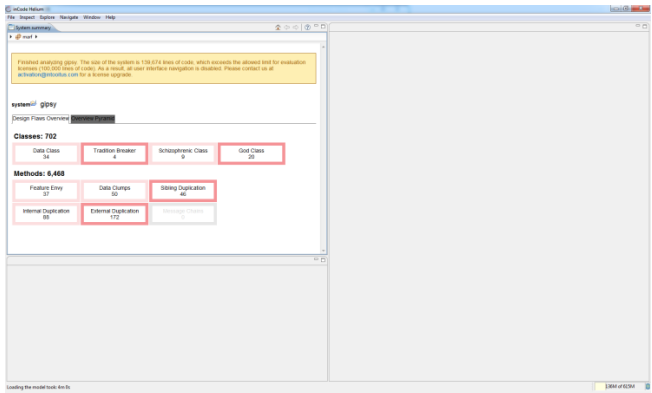

Figure 44: InCode estimates for GIPSY

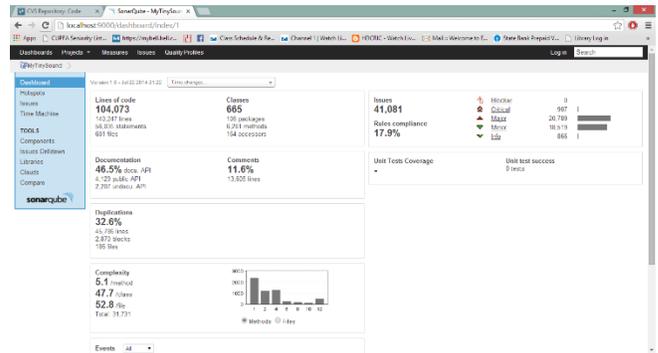

Figure 46: SonarQube estimates for GIPSY

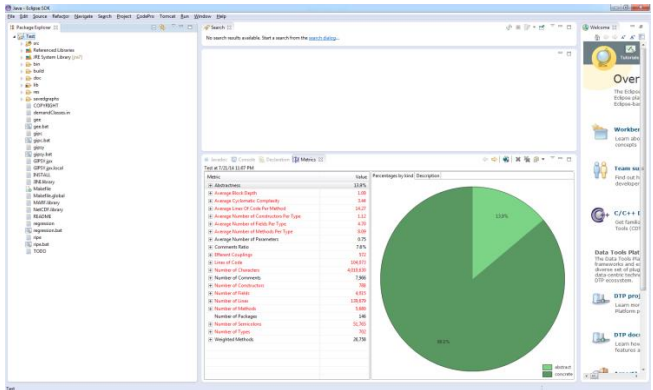

Figure 45: CodePro estimates for GIPSY

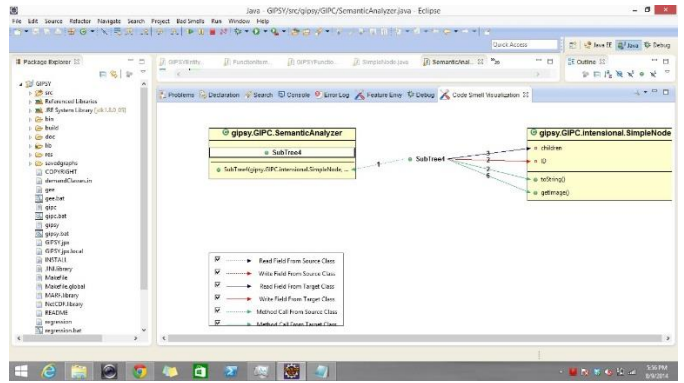